\documentclass{aastex}

\usepackage{rotating}

\newcommand{\eg}{{\it e.g.}}
\newcommand{\ie}{{\it i.e.}}

\newcommand{\etal}{{\it et~al.}}

\newcommand{\dif}{\mathrm{d}}

\newcommand{\aeiH}{(a,e,i,H)}

\newcommand{\au}{\,\mathrm{au}}
\newcommand{\km}{\,\mathrm{km}}

\newcommand{\meter}{\,\mathrm{m}}

\newcommand{\um}{\,\mu \mathrm{m}}

\newcommand{\Myr}{\,\mathrm{Myr}}
\newcommand{\days}{\,\mathrm{d}}

\newcommand{\second}{\,\mathrm{s}}
\newcommand{\mags}{\,\mathrm{mag}}

\newcommand{\gps}{\ensuremath{g_{\rm P1}}}
\newcommand{\rps}{\ensuremath{r_{\rm P1}}}
\newcommand{\ips}{\ensuremath{i_{\rm P1}}}
\newcommand{\zps}{\ensuremath{z_{\rm P1}}}
\newcommand{\yps}{\ensuremath{y_{\rm P1}}}
\newcommand{\wps}{\ensuremath{w_{\rm P1}}}

\newcommand{\PS}{\protect \hbox {Pan-STARRS}}
\newcommand{\PSone}{\protect \hbox {Pan-STARRS1}}

\newcommand{\mps}{\meter\,\second^{-1}}
\newcommand{\magperday}{\mags\,\days^{-1}}

\newcommand{\digest}{{\tt digest2}}

\newcommand{\bottkehnoughtfitted}{23.26}
\newcommand{\bottkehnoughtfittederr}{0.02}
\newcommand{\betafitted}{0.573}
\newcommand{\betafittederr}{0.002}
\newcommand{\vnought}{18.5}

\newcommand{\tcd}{P1010ae}
\newcommand{\deltammin}{11.0}
\newcommand{\deltammax}{12.4}

\newcommand{\bigau}{\,\mathrm{AU}}

\begin{document}

\title{
Observational Constraints on the Catastrophic Disruption Rate of Small Main Belt Asteroids}

\author{
Larry Denneau\altaffilmark{1,2} (denneau@ifa.hawaii.edu),
Robert Jedicke\altaffilmark{1},
Alan Fitzsimmons\altaffilmark{2},
Henry Hsieh\altaffilmark{3,1},
Jan Kleyna\altaffilmark{1},
Mikael Granvik\altaffilmark{4},
Marco Micheli\altaffilmark{5,1},
T. Spahr\altaffilmark{6},
Peter Vere{\v s}\altaffilmark{1},
Richard Wainscoat\altaffilmark{1},
W. S. Burgett\altaffilmark{1}, K. C. Chambers\altaffilmark{1}, P. W. Draper\altaffilmark{7}, H. Flewelling\altaffilmark{1}, M. E. Huber\altaffilmark{1}, N. Kaiser\altaffilmark{1}, J. S. Morgan\altaffilmark{1}, and J. L. Tonry\altaffilmark{1}
}

\slugcomment{61 Pages, 10 Figures, 3 Tables}

\altaffiltext{1}{Institute for Astronomy, University of Hawaii at Manoa, Honolulu, HI 96822, USA}
\altaffiltext{2}{Astrophysics Research Centre, School of Mathematics and Physics, Queen's University Belfast, Belfast, BT7 1NN, UK}
\altaffiltext{3}{Institute for Astronomy and Astrophysics, Academia Sinica, Taipei 10617, Taiwan}
\altaffiltext{4}{Department of Physics, P.O. Box 64, 00014 University of Helsinki, Finland, and Finnish Geodetic Institute, P.O. Box 15, 02430 Masala, Finland}
\altaffiltext{5}{ESA NEO Coordination Centre, Frascati (RM), Italy}
\altaffiltext{6}{Minor Planet Center, Cambridge, MA}
\altaffiltext{7}{Department of Physics, Durham University, South Road, Durham DH1 3LE, UK}

%
%

\shorttitle{Catastrophic Disruption Rate of Main Belt Asteroids}
\shortauthors{Denneau \etal}

\begin{abstract}

We have calculated 90\% confidence limits on the steady-state rate of
catastrophic disruptions of main belt asteroids
in terms of the absolute magnitude at which one catastrophic
disruption occurs per year ($H_0^{CL}$) as a function of the
post-disruption increase in brightness ($\Delta m$) and subsequent
brightness decay rate ($\tau$).  The confidence limits were calculated
using the brightest unknown main belt asteroid ($V=\vnought$) detected
with the Pan-STARRS1 (\PSone) telescope.  We measured the \PSone's catastrophic disruption
detection efficiency over a 453-day interval using the Pan-STARRS
moving object processing system (MOPS) and a simple model for the catastrophic disruption
event's photometric behavior in a small aperture centered on the catastrophic disruption
event. We then calculated the $H^{CL}_0$ contours in the ranges from
$0.5\mags < \Delta m < 20\mags$ and $0.001\magperday < \tau <
10\magperday$ encompassing measured values from known cratering and
disruption events and our model's predictions.  Our simplistic catastrophic disruption
model suggests that $\Delta m \sim 20\mags$ and $0.01\magperday \la
\tau \la 0.1\magperday$ which would imply that $H_0 \ga 28$ ---
strongly inconsistent with $H_{0,B2005} =
\bottkehnoughtfitted\pm\bottkehnoughtfittederr$ predicted by
\citet{Bottke2005} using purely collisional models.  However, if we assume that $H_0 = H_{0,B2005}$
our results constrain $\deltammin\mags \la \Delta m \la
\deltammax\mags$, inconsistent with our simplistic
impact-generated catastrophic disruption model.  
We postulate that the solution to the discrepancy is that $>99\%$ of main belt catastrophic disruptions in the size range to which this study was sensitive ($\sim 100$~m) are not impact-generated, but are instead due to fainter rotational breakups, of which the recent discoveries of disrupted asteroids P/2013 P5 and P/2013 R3 are probable examples.
We estimate that current
and upcoming asteroid surveys may discover up to 10 catastrophic disruptions/year
brighter than $V = \vnought$.

\end{abstract}

\keywords{Surveys:\PS; Asteroids; Disruptions, Collisions}

\maketitle

\section{Introduction}
\label{s.Introduction}

The ever-increasing survey discovery rate of asteroids and comets has
revealed a Solar System teeming with activity. We now understand the
Solar System in terms of inter-related, evolving small body
populations whose orbital element and size-frequency distributions
(SFDs) are sculpted by gravitational interactions with the giant
planets, and by other forces such as the Yarkovsky and YORP effects
\citep{Bottke2006}. These interactions can scatter small bodies to
different regions of the Solar System or eject them entirely.

The main belt SFD has been shaped by collisions between asteroids,
most effectively by `catastrophic' collisions that disrupt an asteroid
into a `family' containing thousands of fragments. \citet{Parker2008}
estimate that at least 35\% of large asteroids ($H < 13$) and 50\% of
smaller bodies are members of a collisional family. Other mechanisms
for SFD modification include but are not limited to: rotational
breakup, sublimation of water ice and other volatiles, thermal
instability, and electrostatic forces. A new class of `active
asteroids' \citep[\eg][]{Hsieh2006,Jewitt2012} is comprised of
dynamically-ordinary main belt objects (expected to be inert) that
display cometary behavior for weeks or even years due to these
mechanisms.

A catastrophic disruption is conventionally defined as a breakup of
an asteroid that leaves no fragment larger than half the
original mass \citep{Greenberg1978}.  Traditionally
this has been understood to occur via a collision between a
parent body and a smaller projectile body.  Current understanding of
catastrophic disruptions is limited by the rarity of these events and
by a preference toward investigation of impacts between
kilometer-scale and larger objects that result in the creation of
asteroid families \citep[\eg][]{Michel2002}.  These studies have
focused on the macroscopic properties of asteroid collisions with less
emphasis on the microscopic aspects of the resulting dust production
and dispersion. Furthermore, while cometary dust behavior has been
studied for many decades, building upon models by \citet{Finson1968}
that describe the orbits of dust grains using radiation pressure and
solar gravity, the behavior of dust generated in asteroidal outburst events
has only recently begun to be studied in detail
\citep[\eg][]{Agarwal2013,Bodewits2011,Stevenson2012b}.

\citet{Bottke2005} employed collisional-cascade simulations
\citep[CoDDEM; also \eg][]{Durda1998,OBrien2003} to predict a main
belt steady-state impact-generated catastrophic disruption rate of about one event per
year for objects of $100\meter$ diameter\footnote{\label{footnote.albedo}The conversion of
  $H$ to diameter depends on an assumed geometric
  albedo. \citet{Pravec2012} measured geometric albedos $p_V$ to be
  $\sim$0.057 for C/G/B/F/P/D type asteroids and $\sim$0.197 for S/A/L
  types. We have assumed an average geometric albedos of $p_V=0.11$ so
  that $H=23$ corresponds to $100\meter$ diameter and $H=18$
  corresponds to $1\km$ diameter.}.  In these simulations, the main
belt is divided into logarithmic size bins, and the population within
each bin evolves over time according to the number of objects lost due
to \eg\ disruption, cratering events, and dynamic depletion that are,
in turn, gained by bins corresponding to smaller asteroid sizes.  They
incorporate estimated rates for the dynamical depletion of the main
belt in the early Solar System due to perturbations from planetary
embryos and a newly-formed Jupiter, and then run their simulations for
the age of the Solar System to fit the current main belt SFD.  Their
models are also subject to additional constraints including the number
of known asteroid families from large ($D > 100\km$) parent bodies,
asteroid (4) Vesta's single large impact crater, and the lunar and
terrestrial impactor flux.

The capabilities of modern asteroid surveys now allow the
serendipitous detection of these disruptions. Surveys such as
Pan-STARRS \citep[\eg][]{Kaiser2002,Hodapp2004} and the Catalina Sky
Survey \citep[CSS; \eg][]{Larson1998,Larson2007} have improved in
capability to routinely detect active objects in the main belt.  When
a survey is well-characterized \citep{Jedicke2002} it is possible to
measure or set limits on disruption rates that constrain the behavior
of large-scale main belt collisional models
\citep[\eg][]{Walsh2009,Bottke2005}.

It remains difficult to observationally identify the cause of activity
in main belt objects. All of the aforementioned mass loss mechanisms
produce clouds of dust or ice grains causing a comet-like appearance
and often a dramatic increase in brightness. The range of behaviors
includes; asteroid P/2010 A2, whose tail structure suggests either an
asteroid impact \citep{Snodgrass2010} or YORP-induced rotational
breakup \citep{Agarwal2013}; the long dust tails of main belt comets
like 133P/Elst-Pizarro \citep{Hsieh2009}, compatible with embedded
volatiles escaping the parent body after perihelion; and P/2012 F5,
activated due to an impact \citep{Stevenson2012b}. At the extreme end
is the 2007 outburst of comet 17P/Holmes, whose apparent brightness
increased by 15 magnitudes (from +17 to +2) in 42 hours
\citep{Stevenson2012a}. The cause of the outburst is unknown but a
similar outburst observed during its discovery apparition in 1892
implies an origin related to its inherent nuclear structure and
perihelion passage. We include comet 17P/Holmes in our examples even
though it is not an asteroid because it illustrates that extreme
changes in flux are possible ($\Delta m \sim 20$) and how a bright
outburst might appear at main belt distances.  Two new members were
added to the active main-belt asteroid lineup as of October 2013:
P/2013~P5 and P/2013~R3, whose morphologies are suggestive of
disruption events \citep{Jewitt2013b,Jewitt2014}.

Despite the increasing numbers of known active main belt objects there
has been no verified identification of a collisional catastrophic
disruption.  We use data from the \PSone\ all-sky survey and a simple
model for a disruption to set an upper limit on the collisional 
catastrophic disruption rate of main belt asteroids.

\section{Method}
\label{s.Method}

\subsection{Disruption model}
\label{ss.DisruptionModel}

Our simple model (fig.~\ref{fig.modeldisruption}) for a catastrophic
disruption makes no attempt to incorporate the dynamics and
interactions of the post-disruption fragments.  We simply assume that
at time $t=0$ a parent body of diameter $D$ is catastrophically
disrupted into a spherical, homogeneous cloud of particles of diameter
$d$ that expands at speed $v$.  Our photometric aperture with diameter
$a$ when projected to the location of the disruption views a
`cylindrical plug' through the spherical cloud, and our model approximates
the reflected light contributed by particles inside the volume of the plug.  The change in
apparent magnitude as a function of time since the catastrophic
disruption can be approximated with:
\begin{equation}
\Delta m(t) = 
\left\{
\begin{array}{l r l}
   - 2.5 \, \log_{10} \Big[{ 4 v^2 \over D^2 } \, t^2                                  \Big] &           0<&t<t_{thick} \\
   - 2.5 \, \log_{10} \Big[{D \over d}                                                 \Big] & t_{thick} < &t \le t_{aperture} \\
   - 2.5 \, \log_{10} \Big[{ { D_0 \over d } \Big[ 1-\frac{\left(t^2 v^2-r_a^2\right)^{3/2}}{t^3 v^3} \Big]} \Big] &             &t > t_{aperture} \\\end{array}
\right.
\label{eq.modeldeltam}
\end{equation}
where $t_{thick}$ is the time span during which the dust cloud remains
optically thick and $t_{aperture}$ is the time during which the dust
cloud is smaller than the projected photometric aperture, $r_a$ is the radius
of the projected measurement aperture at the site of the disruption, and
$R(t)$ is the radius of the dust cloud as a function of time. The appendix (\S \ref{sec.appendix})
shows the derivation of this approximation for the magnitude behavior of a
catastrophic disruption. 

We illustrate our model using a hypothetical impact-generated catastrophic disruption
of a $100\meter$ diameter parent body; \citet{Bottke2005}
predicted that a main belt object of this size is catastrophically
disrupted once/year and because we expect that \PSone\ could detect
this type of disruption.
The simplification that the entire parent body is transformed into
particles of a single diameter is justified by the $n(d) \sim d^{-3.5}$ SFD power-law approximation 
by \citet{Jewitt2009B}
describing the differential number of collisional fragments
as a function of diameter.  The differential
reflected surface area $dA$ contributed by particles of diameter $D$ to $D+dD$ goes as 
\begin{equation}
dA = {\pi \over 4} D^2 D^{-3.5} dD
\label{eq.da}
\end{equation}
and the cumulative normalized contribution $A_{cumul}(D)$ in reflected surface area for particles
smaller than size $d$ is then
\begin{equation}
A_{cumul}(D) = K \Big[D_{min}^{-0.5} - D^{-0.5}\Big]
\label{eq.acumul}
\end{equation}
where $K$ is a normalization constant that produces $1$ for large $D$,
so that $A_{cumul}(D)$ represents the fraction of total reflecting
area contributed by particles of diameter $D$ or smaller.
(eqs. \ref{eq.da} and \ref{eq.acumul} use capital $D$ for readability; 
hereafter, we use $D$ to refer to parent body diameter and $d$ to particle diameter).
We assume a minimum particle size of $1 \um$; below this size, particles
are rapidly swept away by solar radiation pressure and are inefficient
scatterers at the nominal $0.6\um$ bandpass of our measuring system \citep{Fink2012}. Under this scenario, a $100\meter$ diameter parent body that suffers a
barely-catastrophic disruption leaving a single $80\meter$ diameter remnant,
$\sim$90\% of the fragments' reflected light derives from particles
$<100\um$ diameter and $\sim$68\% from those $<10\um$.
The projectile asteroid's mass can be ignored relative to the
parent body because it will typically have about 1/10 the
diameter of the parent body but only 1/1000
of the total mass in the collision \citep{Davis1989}.  Thus, without introducing much
error, we can set an upper bound on the reflected light (and maximize
the detection efficiency) by assuming 100\% conversion of the
parent body into $d=1\um$ diameter particles. For our hypothetical
$100\meter$ body, this results in a brightness increase of +20 magnitudes.
While 100\% conversion represents an upper limit, if as little as 20\% of the 
parent body is converted to $1\um$ particles, the resulting dust cloud would still produce
a brightness increase of over 18 magnitudes.

We assume a fixed photometric measurement aperture of $3.0\arcsec$ to
mimic the detection of a catastrophic disruption by a survey telescope
employing automatic identification of point sources (like \PSone).
This aperture corresponds to a spatial diameter of $a\sim3,000\km$ at
the disruption assuming a typical main belt opposition distance of
$1.5\bigau$ from Earth.

The expanding dust cloud's photometric behavior in the aperture for
$t>t_{aperture}$ was simply modeled with a fixed magnitude decay rate
($\tau$) in the range from 0.001 to 10 $\magperday$.  This range
incorporates essentially the entire range of decay rates predicted by
our simplistic catastrophic disruption model (eq.~\ref{eq.modeldeltam}) when we assume a
representative dust expansion rate of $v=1\mps$. This rate, in turn,
is roughly the geometric mean of the range of ejecta speeds
($\sim0.06\mps$ to $\sim100\mps$) required to model dust ejected from
the $\sim100\meter$ diameter asteroid P/2010~A2 estimated by \citep{Bodewits2011}
in their comparison of ejecta from a cratering event in asteroid (596) Scheila in late 2010.
Thus, our model implicitly incorporates a range of ejecta speeds and
any size-dependence on the progenitor by examining a wide range of
magnitude decay rates.

Summarizing our representative case, prior to collision the to-be disrupted
$100\meter$ diameter object is too faint to be detected by the
system. Immediately following the collision the cross-sectional area
of dust-dominated fragments begins increasing rapidly such that the
disruption reflects enough light to become visible but remains
morphologically indistinguishable from a stellar point source.  The
expanding dust cloud of $1\um$ diameter dust particles remains
optically thick for $\sim6\days$ following disruption as it expands to
a radius of $\sim1,000\km$ and produces a $\sim10^8$ increase in flux
($\Delta m=+20$ from eq.~\ref{eq.modeldeltam}).  There is no further change in
the aperture flux until the dust cloud extends beyond the projected 
measurement aperture. The flux decreases as the dust
optically thins and as the cloud expands such that a smaller portion
of its volume occupies the measurement aperture. The disruption model incorporates three phases: 1) an immediate
ramp-up in brightness during which the disruption becomes detectable
due to rapid expansion of the optically thick dust cloud; 2) a period of
nearly-constant brightness during which the dust cloud remains
optically thick but is expanding beyond the measurement aperture, or has become optically thin but remains completely within the measurement aperture; and
3) a decay period when the disruption is thinning optically and
continuing its expansion well beyond the aperture.  Fig. \ref{fig.mag} illustrates the change in brightness over
time for our representative disruption using various dust expansion velocities.

Motivated by this simple model and its analytical realization in
eq.~\ref{eq.modeldeltam}, and understanding its limited fidelity, we further
simplify our model of the catastrophic disruption as an
instantaneous increase in brightness (decrease in magnitude, $\Delta
m$) followed by a linear decrease in brightness (increase in
magnitude) at a rate $\tau$.  The ramp-up time in brightness due to
disruption and the time spent in the `plateau' are short enough
relative to the decay time and the survey observing cadence that they
can be omitted from the disruption model.  Furthermore, the magnitude
decay rate in our analytical model changes by only a factor of 10 from
20 to 200~days without even accounting for gravitational
reaccummulation of the fragments (not relevant for 100 m bodies \citep{Jewitt2014}), solar radiation pressure, or any other physics, so that our
constant-decay model can span the predicted decay rates.  Thus, the
apparent magnitude at time $t$ after a disruption that occurs at time
$t_{CD}$ is then
\begin{equation}
m_{CD}(t) = m_0 - \Delta m_{CD} + \tau \; ( t - t_{CD} )
\label{eq.m_CD}
\end{equation}
where $m_0$ is the parent body's undisrupted apparent magnitude.  We
will calculate the confidence limit on the catastrophic disruption
rate, \ie\ the largest asteroid that can be disrupting at the rate of
one per year, for $0<\Delta m<20$ and decay rates $\tau$ in the range from
$10^{-3}$ to $10\mags\days^{-1}$ as suggested by our analytical model
and the known set of active asteroids. Fig. \ref{fig.dmag} shows the
{\it change} in magnitude decay over time of our representative disruption.
For a range of dust dispersion velocities from $0.5$ to $5.0\mps$
we expect to see a magnitude decay rate between $0.01$ and $0.1\magperday$,
except for case involving a rapidly expanding dust cloud, \eg\ $5\mps$. 
In this case the decay in brightness
can be quite rapid, several tenths of a magnitude per day, once the optically 
thick phase is over or the cloud has expanded beyond the projected measurement aperture.

\subsection{Catastrophic disruption rate limit}

The expected number of {\it new} detected
catastrophic disruptions in a survey time range $[t_{min},t_{max}]$
with apparent magnitude $V$ brighter that some arbitrary brightness limit $V_{CD}$ and a magnitude profile described by $\Delta m$ and $\tau$ is
\begin{eqnarray}
\nonumber
N_{CD}(\Delta m,\tau;V_{CD}) &=& 
\end{eqnarray}
\begin{eqnarray}
\int\limits_{t_{min}}^{t_{max}} dt \int d\vec x \int dH \;
       \epsilon(\vec x,H,t;\Delta m,\tau) \; f(\vec x,H) \; n(\vec x,H) 
          \; h[(V_{CD}-V(\vec x,H,t;\Delta m,\tau)]
\label{eq.ncd}
\end{eqnarray}
where $\vec x$ represents the six orbital elements, $(a, e, i, \Omega,
\omega, M)$, the semi-major axis, eccentricity, inclination, longitude
of the ascending node, argument of perihelion, and mean anomaly
respectively; $H$ represents the pre-disruption absolute magnitude;
$\epsilon$ is the system efficiency at detecting the catastrophic
disruption; $f(\vec x,H)$ is the fraction of objects disrupted per
unit time; $n(\vec x,H)$ is the number density of objects; and $h(z)$
represents the Heaviside function with $h=0$ for $z<0$ and $h=1$ for
$z\ge0$ that enforces the apparent magnitude of the catastrophic disruption to be $\le
V_{CD}$.  We will select a $V_{CD}$ such that there are no
known catastrophic disruptions brighter than this threshold so
that we can compute an upper limit to the catastrophic disruption
rate.

To compute this limit, and for comparison with theoretical modeling, the
evaluation of eq.~\ref{eq.ncd} requires us to combine the main belt
number density and disruption fraction into an orbit-independent
catastrophic disruption frequency: $\phi(H) \approx f(\vec x,H) \,
n(\vec x,H)$ with the form $\phi(H)=10^{\beta(H-H_0)}$.  \ie\ we
assume that the main belt orbital element distribution is
$H$-independent.  The functional form is motivated by the power-law
relationship for the expected frequency of disruption events for main
belt asteroids as determined by the CoDDEM simulations by
\citet{Bottke2005} over the 4.6~Gyr lifetime of the Solar System.

We converted$^{\ref{footnote.albedo}}$ the CoDDEM simulations'
\citep{Bottke2005} diameter-dependent disruption rate prediction to
absolute magnitude (fig. \ref{fig.bottke}), fit it to a power-law of
the form $f(H) = 10^{\beta(H-H_0)}$, and found $H_{0,Bottke} =
\bottkehnoughtfitted\pm\bottkehnoughtfittederr$ and $\beta =
\betafitted\pm\betafittederr$.  We adopt the power law slope $\beta$
into our formulation of the catastrophic disruption rate limit, described below, and are primarily interested
in observationally determining $H_0^{CL}$, the smallest absolute magnitude (largest diameter) at
which one disruption is occurring per year.
The uncertainties on $H_{0,Bottke}$ and $\beta$ were obtained by
computing the root-mean-square residuals of linear fits to the
data in fig. \ref{fig.bottke} using $2\mags$-wide sliding windows from $9 <
H < 29$ at 1-mag intervals.

The catastrophic disruption detection efficiency calculation is discussed in detail below (\S\ref{s.psonedeteff})
where we will subsume the efficiency with the requirement that the catastrophic disruption
have apparent magnitude $V<V_{CD}$ into a single orbit element and
time-averaged function:
\begin{equation}
\bar\epsilon(H;\Delta m,\tau,V_{CD}) \approx \epsilon(\vec x,H,t;\Delta m,\tau) \; h[(V_{CD}-V(\vec x,H,t;\Delta m,\tau)]
\end{equation}
  
We assume disruptions occur independently at a steady rate and
are Poisson-distributed over time.  If we select a brightness
threshold $V_{CD}$ for which there is a single candidate disruption
at this brightness and none brighter in our dataset, we can compute
a 90\% confidence limit (C.L.) on $H_0$, the largest diameter asteroid
at which one catastrophic disruption occurs per year by integrating
over $H$ multiplied by our system detection efficiency and the number of disruptions predicted
by the \citet{Bottke2005} power law, then solving for $H_0$:
\begin{equation}
\label{eq.hnought}
H_0^{CL}(\Delta m, \tau, V_{CD}) = {1\over\beta} \log_{10} 
 \Biggl[  
   {\Delta t \over 3.9} \int dH \;\bar\epsilon(H;\Delta m,\tau,V_{CD}) \; 10^{\beta H}
 \Biggr]
\end{equation}
where $\Delta t = t_{max} - t_{min}$ is the survey's time
duration.

\subsection{The Pan-STARRS1 survey}
\label{ss.PanSTARRS1}

The \PSone\ telescope \citep[\eg][]{Kaiser2002,Kaiser2004,Hodapp2004}
began operations in late 2008 as a composite survey to satisfy the
various science goals of the \PSone\ Science Consortium. At survey
inception $\sim$85\% of the total survey was executed using cadences
suitable for asteroid detection; that is, the telescope obtained at
least two exposures at the same footprint separated by 10 to 30
minutes. The original asteroid discovery program of producing orbits
from pairs of observations over multiple nights \citep{Kubica2007}
was quickly found to be ineffective due to the larger-than-expected
impact of detector gaps and false detections. The \PSone\ asteroid
survey was reworked in 2010 into a single-night discovery mode using
`quads' (4 exposures per footprint separated by $\sim15$~minutes) for
its dedicated Solar System time and to opportunistically utilize the
other survey modes when possible.

Beginning in early 2010 5\% of the survey time was dedicated to
detection of Solar System objects, in particular near earth objects
(NEOs), and this fraction increased to 11\% as of November 2012.  The
Solar System survey uses the wide-band \wps\ filter and reaches $V
\sim 21.5$ \citep{Denneau2013}. The remaining survey time useful for
asteroid detection consists of 1) the all-sky, multiple filter, 3$\pi$
survey, that acquires pairs of exposures in each of \PSone's
$g,r,i,z,y$ filters (from here on called \gps, \rps, \ips, \zps, and
\yps) and 2) the Medium Deep Survey, a `deep drilling' survey that
obtains 8 $\times$ 240-second sequences at ten different fixed
footprints throughout the year.  The 3$\pi$ survey has a single-epoch
point source sensitivity of $V \sim 20.5$ in the $g$ filter and $V
\sim 21$ using $r$ and $i$.

Successive Pan-STARRS1 images at the same footprint are automatically
reduced by the \PSone\ Image Processing Pipeline
\citep[IPP;][]{Magnier2006} and then subtracted pairwise to identify
transient detections. IPP delivers catalogs of transient detections to
the Moving Object Processing System \citep[MOPS;][]{Denneau2013}. MOPS
assembles `tracklets', associations of transient detections across
multiple exposures at the same footprint from the same night that
\emph{may} represent a real asteroid, that typically contain 3 or 4
detections from a `quad'. MOPS also creates tracklets from the Medium
Deep survey's 8-exposure sequences and from pairs of 3$\pi$ exposures.
IPP achieves better than $0.05\mags$ photometric uncertainty for
bright asteroids detected in the subtracted images and $\sim
0.15\arcsec$ astrometric uncertainty \citep{Milani2012} for all
asteroids over the entire sky. \PSone\ has
delivered more than 7 million detections to the IAU Minor Planet
Center (MPC) as of February 2014.

The \PSone\ 3$\pi$ observing cadence is designed to obtain tracklets
at six epochs in each lunation with the \gps, \rps, and \ips filters.
In theory, this cadence should have enabled \PSone\ to perform enough
self-followup of each detected object to allow an orbit determination.
In practice, the combined effects of weather losses and detector gaps
conspired to rarely allow an orbit calculation.  Thus, nearly all
\PSone\ `interesting' asteroids and comets are confirmed with other
observational facilities.

\subsection{Disruption candidate search}
\label{ss.Search}
Our search for main belt catastrophic disruption candidates relies on
the determination by the MPC that a \PSone\ asteroid tracklet cannot
be linked with a known object. We argue that such an unlinked tracklet
observed in opposition with main-belt motion represents a catastrophic
disruption candidate if it is brighter than the limit to which the
main belt population is believed to be complete.

Unlinked tracklets for all observatories are published regularly by
the MPC in its `one-night stand' (ONS) file. The \PSone\ contributions
to the ONS file (fig.~\ref{fig.onslist}) are usually a) NEOs that were
not followed up and are now `lost'; b) faint main-belt asteroids seen
too few times to produce an orbit; or c) false tracklets due to image
artifacts or mis-linkages that were submitted automatically by MOPS.

A \PSone\ tracklet composed of {\it bright} detections in the ONS with
main belt-like rates of motion must therefore be a main belt asteroid
that has escaped detection until now, or an NEO `hiding in plain
sight' (both exceedingly unlikely for bright objects), or a small main
belt asteroid activated into detectability due to a disruption event.
The brightest real, unlinked \PSone\ tracklet satisfying these
conditions over the period 2011-02-21 through 2012-05-19, submitted
with the identifier \tcd, has apparent magnitude $V = \vnought$
(fig.~\ref{fig.onslist} and table~\ref{t.params}).  The detections in
the \tcd\ tracklet have a positional great-circle residual of
$0.04\arcsec$ and photometric variation of $0.07\mags$, consistent
with expectations for a main belt object.  The detections show no sign
of `activity' as they are morphologically consistent with the
point-spread functions (PSF) of nearby flux-matched stars.

Main belt asteroids with apparent magnitude $V<16.5$ are nearly
certain (fig.~\ref{fig.hcompleteness}) to have absolute magnitudes
less than the completeness limit of $H=15$ reported by
\citet{Gladman2009}.  We suggest that the main belt is complete to a
smaller size limit corresponding to $H\sim17.5$ because our disruption
candidate, \tcd\ with $V = \vnought$, is the brightest unlinked
(unknown) \PSone\ tracklet in the MPC ONS file.  Figure
\ref{fig.hcompleteness} shows the probability that a main belt asteroid
detected near opposition has $H<H_{complete}$ as a function of its
apparent magnitude for $H_{complete}$ = 15, 16 and 17.  We see that an
asteroid near opposition with $V = \vnought$ is virtually certain to
be known if the main belt is complete to $H_{complete} = 17.5$.  
We base this statement on the fact that millions of asteroids have
been observed in opposition in \PSone, and if the main belt were not
complete to $H_{complete}=17.5$, there should be many unlinked
ONS tracklets with apparent magnitude brighter than $V=\vnought$ when there are none (there are only a handful brighter than $V=19$).  But even if
\tcd\ is in fact only a main belt asteroid, and not the remnant of a
catastrophic disruption, our computed catastrophic disruption rate limit based on this brightness threshold remains valid because there
are still no detected catastrophic disruptions {\it brighter} than
$V=\vnought$ and we have simply overestimated the upper limit on the
rate.

We use the MPC \digest\ score\footnote{{\tt digest2} was developed by
  S. Keys, C. Hergenrother, R. McNaught, and D. Asher, and is
  available at {\tt https://code.google.com/p/digest2}.}
\citep{Jedicke1996} to determine whether a single-night tracklet is
likely to be a main belt asteroid.  \digest\ generates a set of
virtual orbits that are statistically compatible with the input
tracklet (\eg\ \citet{Virtanen2001,Muinonen2006}) to compute a
pseudo-probability that a tracklet belongs to a particular asteroid
population.  It is used by asteroid surveys to select tracklets for
immediate followup (typically NEOs) that have not been associated with
known objects.  In this study we selected ONS tracklets that have
\digest\ scores of $>25$ for at least one of
the MB1 (inner), MB2 (middle) or MB3 (outer) main belt
classes.\footnote{In \digest, MB1 is defined generally as $2.1 < a <
  2.5$ and $q > 1.67$, MB2 as $2.5 < a < 2.8$, and MB3 as $2.8 < a <
  3.25$, where $a$ is the orbital semi-major axis and $q$ the
  perihelion distance in astronomical units.}
  
To summarize: we use the MPC ONS database and the {\tt digest2}
computation to select an unknown asteroid with main belt 
motion with a brightness that suggests the asteroid has a size
in the complete population.  Since it is unknown, we consider it
a candidate catastrophic disruption and choose its apparent magnitude
as $V_{CD}$ for eq. \ref{eq.hnought}.

Our calculation of the main belt catastrophic disruption rate limit
requires that we identify candidate disruptions that occur {\it in}
our observation window. A possible source of confusion is disruption(s)
that occur prior to the observing window but remain visible.  
The ability of the current asteroid surveys to repeatedly
image the main belt near opposition in single or adjacent lunations
suggests that a bright, slowly-decaying disruption would be observed
on multiple nights and linked by the MPC into an asteroid orbit;
\ie\ it would not appear in the ONS and is unlikely to be one of our
candidate disruptions.  Conceptually this situation could lead to 
slow-decaying disruptions `hiding' undetected in \PSone\ data, leading
to an erroneous computation of the limit.  However, recent discoveries of faint
($m > 20$) active asteroids identified automatically by current all-sky surveys suggest that a slow-decaying disruption
brighter than our candidate object ($m < 18.4$) would eventually have been detected by
virtue of its comet-like appearance by at least one of the surveys and would therefore be confirmed
as an actual disruption having a disruption date outside our survey window.  That no such event was observed
leaves us confident that our single ONS candidate
with magnitude $V=\vnought$ represents the brightest possible catastrophic
disruption in our observation window.  Since our simple model allows for
a brightness increase of up to 20 magnitudes post-disruption, an additional
consideration is a slow-decaying disruption that actually saturates
the detectors of current surveys, rendering them `undetected' by automatic 
source-finding software.  We argue that an event such as this occurring in the main belt 
would have an obvious cometary signature and would similarly
result in a confirmed disruption.

Alternatively, bright, rapidly-decaying
disruptions, \eg\ $>0.1\mags\days^{-1}$, are unlikely to be observed
again by current all-sky surveys because they fall below the systems's detection thresholds
within weeks. Thus a rapidly-decaying disruption occurring
prior to but detectable within the observation window would have to
occur close enough in time to essentially be considered `in' the
window.
Fig. \ref{fig.onsdecisiontree}
illustrates this reasoning schematically. Note that we are not strongly
concerned with whether our candidate ONS tracklet is truly a disruption --
it may simply be an unknown (at time of observation) main belt asteroid. 
While our analysis of main belt completeness suggests that
the number of unknown main belt asteroids that can have brightness $V=\vnought$ in opposition
is exceedingly small, this would not affect
the validity of our computed catastrophic disruption rate limit, as
we are still stating that there are no catastrophic disruptions {\it brighter}
than our candidate.

\subsection{\PSone\ detection efficiency for catastrophic disruptions in the main belt\label{s.psonedeteff}}

The \PSone\ MOPS software allows us to measure the system's efficiency
(used in eq.~\ref{eq.hnought}) by injecting synthetic detections into
the processing stream to determine how many objects would be detected.
Faint objects can be missed simply because they are not brighter than
the system's limiting magnitude but even bright objects can go
undetected if they saturate the detector, fall in a gap between the
detectors, or are embedded in a bright star's PSF.

For this study, we were not interested in the detection of asteroids
\emph{per se} but the aftermath of the catastrophic disruption of an
asteroid according to our model. Intuitively we expect that
disruptions with larger changes in brightness ($\Delta m$) and longer
decay times ($\tau$) should be easier to detect, as large changes in
magnitude mean that disrupted small asteroids will become bright
enough to be detected, and long decay times give the survey a longer
window in which the disruption is detectable. At the same time, the
very brightest disruptions will saturate the system's detector,
resulting in a window of detectability where the disruption is
brighter than the system sensitivity limit yet below the saturation
limit.

We employed MOPS using \PSone\ telescope pointings over the 453-day
interval from $t_{min}=$ 2011-02-21 through $t_{max}=$ 2012-05-19 and
a realistic distribution of 1.5M main belt objects from a synthetic
Solar System model \citep[S3M;][]{Grav2011}. We exploited our
assumption of the separability of the main belt objects's orbit
distribution from $H$ and assigned all the 1.5M main belt objects
(candidate parent bodies) an absolute magnitude of $H_{ref} = 18$ (the
actual value does not matter). The objects were propagated to the
times of observation of all the \PSone\ exposures and those objects
that were located in any field of view (FOV) were stored in a database
as `in-field detections'.  \ie\ we stored them in the database
regardless of their apparent magnitude at the time they were in the
field.  Each object in that database has entries that include its time
of observation, topocentric position, apparent magnitude $V_{ref}$,
and signal-to-noise ratio.  Approximately 1M of the 1.5M synthetic
asteroids appear in the $\sim$60,000 survey fields, an average of
$\sim100$ per exposure.  The advantage of this technique is that we
can assign any other absolute magnitude $H^\prime = H_{ref} + \Delta
H$ to an object and its apparent magnitude in any field $i$ will
simply be $V_i^\prime = V_{i,ref} + \Delta H$.

We randomly selected $N=1,000$ of the objects to undergo a
catastrophic disruption for each $\Delta m$, $\tau$, and $H^\prime$
combination in a grid of $0.5\mags$ steps for $0.5\mags < \Delta m <
20\mags$, 
0.08 dex for $0.001\magperday < \tau
< 10\magperday$, and $0.5\mags$ steps for $14\mags < H^\prime <
30\mags$.  The catastrophic disruptions occurred at random times distributed uniformly over
the simulation interval and their subsequent apparent magnitude
followed eq.~\ref{eq.m_CD}.  We then applied the filter- and
survey-dependent \PSone\ tracklet identification efficiency
\citep{Denneau2013} to each of the observed synthetic catastrophic
disruptions.  The \PSone\ fill-factor of $\sim0.75$ results in a
tracklet identification efficiency of $\sim$0.7 for the creation of 3-
or 4-detection tracklets from the four exposures obtained in the
\wps\ filter Solar System survey.  The \gps, \rps\ and
\ips\ observations have a tracklet identification efficiency of
$\sim0.5$ because they are acquired using pairs of exposures to create
2-detection tracklets.  We set the detector's saturation limit at the
brightest $V$-band magnitude reported for an asteroid by \PSone, (92)
Undina with apparent magnitude $\ips\ = 13.3$ or $V=13.7$.  The
catastrophic disruption detection efficiency is then $\epsilon(\Delta
m, \tau, H, V_{CD}) = n(\Delta m, \tau, H) / N(\Delta m, \tau, H)$, where $n$
is the number of detected synthetic catastrophic disruptions in a
\PSone\ field brighter than our candidate disruption \tcd\ $(V =
\vnought)$ but not saturating the detector and $N$ is the number of
generated disruptions ($1,000$).

The calculated \PSone\ detection efficiency (fig.~\ref{fig.hgroup})
for catastrophic disruptions of main belt objects agrees with our
expectations that long-lived and brighter events are easier to detect
(brighter by virtue of parent body size or increase in brightness).
The maximum detection efficiency of about 50\% in each $H$ slice
always corresponds to the slowest magnitude decay rate of
$0.001\magperday$. At the slowest decay rates the disruptions's
aftermath decreases in brightness by only about $0.5\mags$ in the
course of the 453 day duration of the survey data --- so, if it is
detectable, it is detectable the entire time.  The survey duration
corresponds to the synodic period of main belt asteroids with
semi-major axes of $\ga3\bigau$ so that almost every disruption further
from Earth at opposition than about $2\bigau$ will be covered if it is
above the system's limiting magnitude.  On the contrary, about 60\% of
main belt objects have synodic periods $\ga453\days$ so that the
survey's completeness for main belt objects is distance-dependent.
This factor, combined with the system's limiting magnitude and
saturation limit, the survey coverage, and the camera's fill-factor
conspire to result in a maximum catastrophic disruption detection
efficiency of $\sim0.5$.  The pattern of the efficiency contours
shifts to larger $\Delta m$ as the absolute magnitude of the parent
body increases because smaller objects require larger changes in flux
so that they are bright enough to be detected.

\subsection{Catastrophic disruption rate limits}

Our 90\% confidence limit contours on $H_0$ (fig.~\ref{fig.h0})
represent the largest object at a given $(\Delta m, \tau)$ that can be
disrupted in the main belt at the rate of one per year and still be
consistent with the \PSone\ data; \ie\ the actual $H_0$ value must be
greater than the contour value at $(\Delta m, \tau)$.  The contours
are generated by computing $H_0^{CL}$ from eq. \ref{eq.hnought} at
equally-spaced $(\Delta m, \tau)$ grid points.  We draw
lines of constant $H_0$ that indicate what brightness behavior, described
by $(\Delta m, \tau)$, is allowed for a given constant $H_0$ contour to
satisfy the computed limit $H_0^{CL}$ at the $(\Delta m, \tau)$ location.
Because we have computed a limit, a given $H_0$ contour defines a boundary
of behavior, \ie\ `allowed' values for $H_0$ for a given $(\Delta m, \tau)$
lie along the boundary or above and to the left, where objects are smaller. 

Our computed limit space spans
more than 20 absolute magnitudes in our $(\Delta m, \tau)$ parameter
space from $8 < H_0^{CL} < 30$.  
We see that slowly-decaying disruptions with large magnitude changes
with $\Delta m = 15$ and $\tau = 0.001\magperday$) yield
$H_0^{CL}\sim27$, or about $15\meter$ in diameter.  Even though in this case the
parent body is smaller than a house, the disruption of such
a body is bright and long-lived enough to be easily detectable by \PSone.  At
the opposite end, fast-decaying disruptions with small brightness
changes (\eg\ $\Delta m = 0.5$, $\tau = 10\magperday$) would be
detected so rarely by \PSone\ that our method can only constrain $H_0$
to be $>10$ (about $40\km$ diameter).
The computed limit space in itself
does not suggest a particular $H_0^{CL}$; it simply describes
a relationship between $H_0$ and $(\Delta m, \tau)$ and we must
employ additional based on our simple model and other observational data
to narrow down a small region in this space from which we
can select $H_0^{CL}$.

\section{Discussion}
\label{s.Discussion}

Table \ref{t.activeobjects} summarizes the relevant properties of the
recently discovered active (mass-losing) main belt objects and one
near-catastrophic cometary disruption \citep{Reach2010} including our estimates of $\Delta m$
and $\tau$ based on published results. We include those that may be activated by impacts (P/2010~A2, (596)~Scheila and
P/2012~F5~(Gibbs)), rotational breakup (P/2010~A2 again, P/2013~P5 and P/2013~R3) and several others whose activity is probably
produced by other, unknown mechanisms.

Asteroid P/2010~A2 appears twice in fig.~\ref{fig.h0} because it was
first attributed to a impact-generated disruption by \citet{Jewitt2010} who
estimated $\Delta m = +15^{+4}_{-0}$ assuming a $2\meter$ diameter
projectile, but later analysis of precovery observations by
\citep{Jewitt2011} and detailed dust-modeling work by
\citet{Agarwal2013} suggested instead a rotational breakup with
$2.3<\Delta m<5.5$.  Both scenarios allow us to estimate the decay
rates assuming a disruption in early 2009 \citet{Jewitt2010};
$\tau_{collision}$ can be in the range from about $0.055\magperday$ to
$0.070\magperday$ and $\tau_{rotation}$ from about $0.008\magperday$
to $0.020\magperday$ to achieve the apparent nuclear magnitude
measured in HST observations 270 days later \citet{Agarwal2013}.

Asteroid (596)~Scheila ($\sim113\km$
diameter\footnote{\label{footnote.Scheila}{\tt\ http://ssd.jpl.nasa.gov/sbdb.cgi?sstr=596}})
was observed $\sim11\days$ after the impact of a decameter-scale
projectile \citep{Ishiguro2011,Bodewits2011} from which we derive
$\Delta m = 1.0\pm0.1$ and $\tau = 0.087\pm0.007\magperday$.  There
are no reported nuclear observations of asteroid P/2012~F5~(Gibbs)
near the start of activity but we calculate $\Delta m=2.1\pm0.1\mags$
based on its pre-impact absolute magnitude of $17.22\pm0.21$ from
precovery observations \citep{Novakovic2014} and the post-impact
observations of \citet{Stevenson2012b}.

Recent discoveries P/2013~P5 and P/2013~R3 add to an already diverse
assortment of active main belt asteroids. Upon discovery in August 2013 the dust
surrounding P/2013~P5 exhibited rapidly-changing episodic `pinwheel'
behavior. \citep{Jewitt2013b} conclude that this body has been spun
up by radiation torques to the edge of stability. This effects of episodic,
overlapping, mass loss over five months defy simple parameterization using
a single brightness increase and decay rate.

P/2013~R3 was discovered in late 2013 already spit into at least 10 comet-like fragments,
the largest with an effective diameter $\la$ 400~m \citep{Jewitt2014}.  Analysis
of the fragments and surrounding dust suggest a disruption event between February and September 2013.
\citet{Jewitt2014} measured brightness decreases of individual fragments 
ranging from 0.005 to $0.64\magperday$. Averaging these values and extrapolating
to the midpoint of estimated disruption dates provides a crude brightness
increase of +4 magnitudes at the time of disruption.

The uncertainties on our derived $\Delta m$ and $\tau$ values for
individual objects are based purely on the reported uncertainties in
the source observations.  The actual uncertainty must be considerably
larger and is certainly dominated by our incomplete understanding of
the nature of the activity.  All but one of the derived $\tau$ values
(table~\ref{t.activeobjects}) for these objects lies in the range from
$\tau=0.01\magperday$ to $\tau=0.1\magperday$ predicted analytically
from our disruption model.  P/2012~F5~(Gibbs) with
$\tau\sim0.002\magperday$ is anomalously low, perhaps because it was
initially activated by a cratering event leading to sublimation
activity of the newly exposed material.

It is more difficult to constrain $\Delta m$ because none of the aforementioned known
events is thought to represent an collisional catastrophic
disruption.  Our simple model does not incorporate specific collisional features,
so one can argue that rotationally disrupted P/2013~R3 represents the best available model for any
catastrophic disruption even with its modest brightness increase of +4 magnitudes.  
Conversely, for a collisional catastrophic disruption we expect much larger
amounts of dust to be released than from rotational disruption and therefore
a greater increase in magnitude than seen for P/2013~R3. Even if 10\% of a 100~m parent body is converted
to $1\um$ dust, we would see a brightness increase of +17.5 magnitudes; the
initial collisional analysis of P/2010~A2 by \citet{Jewitt2010} corroborates
this.  It is hard to devise a geometric
scenario in an impact event that occludes enough $1\um$ dust to prevent brightness increases
in $+15-20$ magnitude regime.
Thus the aforementioned events provide lower bounds
for $\Delta m$, but are not thought to represent the brightness behavior of an collisional catastrophic disruption. The event with a brightness increase most similar to the
$\Delta m = +20$ predicted by our simple model is the outburst of
comet 17P/Holmes with $\Delta m = +17$, but since this object is not
asteroidal the comparison is questionable.

With no confirmed collisional catastrophic
disruptions and therefore a lack of
observational evidence regarding the dust SFD in a catastrophic disruption, it is reasonable to consider that our assumptions about minimum
dust particle size of $1\um$ and complete conversion of original mass into dust in a collisional
catastrophic disruption may simply be erroneous. If
a typical disruption produced `clumpy' fragments in which only 10\% 
of the mass is converted to dust $100\um$ and larger, the brightness
would increase by only +12.5 magnitudes, largely consistent with \citet{Bottke2005}.
We reject this position based on the following considerations: a) small rubble-pile
asteroids such as (25143) Itokawa may have interiors dominated by fine dust
grains in the micron size regime \citep{Scheeres2012,Tsuchiyama2011};
b) the estimates by \citet{Jewitt2010} of a +15 magnitude brightness increase 
for the sub-catastrophic disruption of P/2010~A2, revised downward after
the dust morphology suggested rotational breakup; and c) the apparent scarcity
of collisional disruptions among recent disruption events.

If we accept our model's $\Delta m =
+20\mags$ and adopt a nominal decay rate at the geometrical mean
between 0.01 and 0.1 $\magperday$, fig.~\ref{fig.h0} suggests that the
largest bodies undergoing catastrophic disruptions at the rate of one
per year have an absolute magnitude no smaller than $H\sim28.7$,
\ie\ no larger than about $7\meter$ in diameter.
Alternatively, we can accept the
$H_0=\bottkehnoughtfitted$ value that we obtained from a fit to the
main belt collisional evolution simulations of \citet{Bottke2005}.
Their results for the resulting main belt SFD are supported by several lines of evidence
that they used to constrain their model (see
\S\ref{s.Introduction}).  With
$H_0=\bottkehnoughtfitted$ and the restricted apparent magnitude decay
rate range of $0.01\magperday \la \tau \la 0.1\magperday$, our results (fig.~\ref{fig.h0}) suggest that $+11.0\mags \la \Delta m \la +12.4\mags$ --- 2 to 4 orders of magnitude smaller in flux than the
$+15\mags$ to $+20\mags$ suggested by our simple disruption model and
impact-scenario modeling of P/2010~A2.  
The naive interpretation of the discrepancy between our $H_0^{CL}$ and
the \citet{Bottke2005} prediction is simply that the main belt catastrophic disruption rate (collision or otherwise) is actually much lower than
they predict. However, while no \emph{impact-generated} catastrophic
disruption has yet been observed in the main belt, ongoing all-sky
surveys have detected several objects undergoing
rotational disruption: P/2010~A2, P/2013~P5 and P/2013~R3.  The prevalence of disruption due to rotation with none due
to collision suggests that rotational disruption may be the dominant
cause of disruption of small asteroids (\eg\ $100\meter$ diameter and
smaller), or that rotational disruptions may occur on longer timescales
than collisional disruptions, resulting in an observational bias. While
our detection efficiency simulations account for slowly-decaying
disruptions, the episodic behavior of P/2013~P5 exposes limitations 
in our simplistic modeling.

Indeed, the \citet{Bottke2005} collisional model did not consider the
possibility of rotational-distribution, and attributed the entire
collisional cascade evolution to impact-generated catastrophic
disruptions, cratering and spallation events.  They then fit their
collision evolution model and resulting main belt size-frequency
distribution to the known population including the smallest objects in
the $100\meter$ diameter size range extrapolated from the NEO
population and cratering statistics.  By ignoring the
rotation-generated catastrophic disruptions their model would over-estimate the size at
which one impact-generated disruption occurs per year.  But it is these impact-generated catastrophic disruptions that would
generate a photometric signature predicted by our simple model.
Thus, it may be possible to reconcile the $H_0=\bottkehnoughtfitted$
predicted by \citet{Bottke2005} with our results by bifurcating the
small-body disruption flux into impact-generated and
rotation-generated components as suggested by \citet{Jacobson2014}.
If impacts play only a small role in generating main belt catastrophic disruptions it could
explain why our study yields a much lower impact-generated catastrophic disruption rate than that predicted
by \citet{Bottke2005} and why contemporary asteroid surveys regularly
identify the aftermath of the rotation-induced disruptions
of main belt asteroids.  Since the collision cascade evolution models incorporate the impact probabilities between main belt asteroids that are probably not too much in error, the implication is that including the rotation-induced catastrophic disruptions in the evolution models will require a modification of the asteroids' strength versus size (the $Q^*_D$ specific disruption energy function in their models, \eg\ \citep{Bottke2005}). These models describe the energies required to disrupt asteroids in both the strength regime (small bodies) and gravitational regime (large bodies), with a minimum unit specific disruption energy $Q^*_D$ occurring between the two regimes near 200~m diameter \citep{Bottke2005}. If rotational breakup is much more common than impact disruption at small sizes, the
intrinsic strength of these bodies may be weaker than predicted by current theories.

\citet{Jacobson2014} suggest that the timescale for YORP-induced
rotational acceleration to the point of disruption for a $100\meter$
diameter main belt asteroid with semi-major axis of $2.5\bigau$ is on
the order of about $0.5\Myr$ while \citet{Farinella1998} estimate that
the impact-generated disruption of the same size object occurs on a
timescale of $\sim200\Myr$.  \ie\ rotation-induced disruption occurs
about $400\times$ more frequently at the asteroid sizes to which
modern surveys are sensitive.  Indeed, rotation-induced
catastrophic disruptions should dominate the detected population of catastrophic disruption
events.

Finally, we can compare the ratio of the number of main belt objects
at our $H^{CL}_0\sim28.7$ (about 7~m diameter) to the number of $H_0\sim23.3$ (100~m) asteroids to determine
whether our measured deficit in the impact-generated catastrophic disruption rate can be recovered by the much more larger rotational catastrophic disruption rate found by \citet{Jacobson2014}.  The SFD is not well measured in this size range
but an accurate value is not important and we simply assume a
canonical dependence on the absolute magnitude proportional to
$10^{0.5\,H}$ \citep{Dohnanyi1969,Dohnanyi1971}.  In this case the
ratio between the $H=28.7$ and the $H=23.3$ populations is a factor of about 500 --- comparable to the ratio between
rotation-induced and impact-generated catastrophic disruptions.  In
other words, our confidence limit and simplistic model may in fact be
consistent with the observations: impact-generated catastrophic disruptions of asteroids in
the $100\meter$ diameter range are extremely rare and the data
suggests with 90\% confidence that the impact-generated catastrophic disruptions have
$H_0^{CL}<28.7$.

Future efforts to improve upon our $H_0^{CL}$ limits should take into
account the following considerations:
\begin{itemize}
\item Our catastrophic disruption model and its implementation as a
  simple linear decay in apparent magnitude is certainly too
  simplistic to represent their actual photometric behavior.  Instead
  of a rapid increase in flux followed by a slow decay, there may be a
  period of rapid decay during which small particles evacuate the
  system due to solar radiation pressure, followed by slower decay as
  the larger particles disperse. This would lead to a shorter period
  of detectability for small (\eg\ $100\meter$ diameter) objects and
  push $H_0^{CL}$ toward larger diameters (smaller $H_0^{CL}$).

\item We have assumed that the central region of a main belt
  catastrophic disruption as viewed from Earth will be detectable with
  the automated \PSone\ IPP software.  This requires that there be a
  central `nuclear condensation' in the photometric aperture.  These
  assumptions have not been tested, though we know that \PSone's
  detection efficiency for `fuzzy' objects is not zero because the
  system has discovered $\sim$40 comets based on their
  almost-but-not-stellar PSFs.  A detection system that could
  automatically detect and characterize extended sources would have a
  much higher detection efficiency for catastrophic disruptions.

\item Main belt collisional evolution models
  \citep[\eg][]{Bottke2005,OBrien2003} are currently only constrained
  by the measured main belt SFD for asteroids $\ga1\km$ diameter so
  the models may overestimate the number of objects in the small size
  regime that contribute to the catastrophic disruption `signal'.  If
  there are fewer small main belt asteroids than predicted then the
  rate of detectable catastrophic disruptions might be less than
  expected.  \citet{Willman2010} proposed that there is a deficit in
  the number of small main belt objects based on space weathering
  gardening times and \citep{Chapman2002} make the same proposition to
  explain the unexpectedly low number of craters on asteroid
  (433)~Eros.  \citet{Obrien2009} however argue that this deficit is
  due to seismic shaking erasing the small craters.

\item Our ranges for $\Delta m$ and $\tau$ may not be representative
  of the behavior of catastrophic disruptions. Fig.~\ref{fig.h0} shows
  that there is much more than a factor of 10 difference in the
  measured magnitude decay rates between outburst events
  (\eg\ P/2012~F5 and P/2010~A2), and $\Delta m$ is not constrained by
  any observations of impact-generated catastrophic disruptions.

\item There may be a parent body size-dependence that is not captured
  by our catastrophic disruption model. The $\sim113\km$
  diameter$^{\ref{footnote.Scheila}}$ asteroid (596)~Scheila is
  thought to have been struck by a decameter-sized object
  \citep{Jewitt2012} that resulted in the creation of a crater and
  which caused a sudden $\sim 1\mags$ brightness increase that faded
  over the course of one month.  We calculated its brightness decay
  rate to be $0.087\pm0.015\magperday$, the fastest in our sample,
  using an impact date estimated by \citet{Ishiguro2011} and nuclear
  photometric measurements by \citet{Bodewits2011}. Even though this
  event was a cratering event it illustrates that the largest
  surviving fragment after disruption can recapture dust particles
  that would otherwise contribute to an expanding dust cloud.

\item The assumption here and in main belt collisional evolution
  models of the separability of the main belt objects's $(a, e, i)$
  from their absolute magnitudes $H$ is certainly not correct in
  detail.  For instance, main belt families have different SFDs
  \citep[\eg][]{Parker2008} from the background population and occupy
  distinct regions in orbit element phase space
  \citep[\eg][]{Zappala1990,Milani1994}.  Thus, there are certainly
  locations in the $\aeiH$ phase space where catastrophic disruptions
  are more or less likely to occur and this will alter their
  detectability by asteroid surveys.

\item The properties of dust evolution after a catastrophic disruption
  has not been investigated and our assumption that all the mass is
  converted into $1\um$ dust particles is clearly an
  oversimplification.  For instance, \citet{Sanchez2013} have shown
  that van der Waals forces between regolith grains can lead to a
  scale dependence in asteroid strength.  The same force would lead to
  `clumpiness' in the dust cloud after a disruption and reduce the
  cross-sectional area of the dust and the observed $\Delta m$.
  
\item The contribution of rotation-induced catastrophic disruptions to
  small-body physical evolution must be understood
  \citep[\eg\ ][]{Jacobson2014}.  Recent events and objects like
  P/2010~A2, P/2013~P5 and P/2013~R3 may help establish the relative
  contribution of rotation- and impact-generated catastrophic disruptions.

\end{itemize}

Looking forward, we can predict the annual number of catastrophic
disruptions that will be detected with $V<\vnought$ using ongoing and
upcoming asteroid surveys (table~\ref{t.ndetectedbysurvey}) assuming
that our single \PSone\ catastrophic disruption candidate is real.
Table~\ref{t.ndetectedbysurvey} provides our conservative estimates
for the `catastrophic disruption survey efficiency' relative to
\PSone\ based on real or anticipated areal sky coverage, limiting
magnitude, cadence, weather losses and observing efficiencies.  For
the purpose of this estimate we define the useful main belt catastrophic disruption search
area as a region $60\arcdeg$ wide in ecliptic longitude
$\times30\arcdeg$ high in ecliptic latitude centered at the opposition
point.  We note that the search for very bright main belt objects like
the immediate aftermath of a catastrophic disruption might actually
benefit from surveying {\it further} from opposition where phase angle
effects reduce an event's apparent brightness so that it does not
saturate the detector.  The catastrophic disruption survey efficiency increases with the
opposition area coverage, limiting magnitude, and per-exposure
detection efficiency. The effect of a more rapid cadence is more
complicated because it can allow for both detection of faster-decaying
events and compensate for area loss on the detector by allowing
multiple opportunities to observe an event. Conservatively, we allow
that multiple coverage of the opposition region per lunation increases
the detection efficiency to 100\% and that re-observations across
lunations provide a factor of 2 improvement.

The opportunity for discovering and studying more catastrophic
disruptions of small main belt asteroids in the next decade is
promising (table~\ref{t.ndetectedbysurvey}) --- if our single
\PSone\ candidate is real.  If it is not real the predicted number
will drop, but it cannot decrease too much without becoming seriously
in conflict with the predictions of the main belt collision models
\citep[\eg][]{Bottke2005,OBrien2003}.  The surprising prediction that
ATLAS \citep{Tonry2011} might detect about a dozen catastrophic
disruptions per year is due, perhaps paradoxically, to its very bright
saturation limit at $V\sim+6$ and its rapid all-sky cadence.  Anything
on the sky brighter than about $V=18$ moving with main belt-like rates
of motion must be something unusual because all the asteroids with
apparent $V<18$ are known.  Thus, ATLAS may detect catastrophic
disruptions of decameter-scale diameter main belt asteroids.
Curiously, LSST's saturation limit of $V = 18$ \citep{Kantor2013}
ensures that it will discover nearly zero disruptions with $V <
\vnought$ but it should excel at the discovery of catastrophic
disruptions by identifying them with their morphological behavior.

It may be possible to identify the remnants of other catastrophic
disruptions via a thorough search of the MPC's dataset beyond the ONS
file. The measured magnitude decay times for P/2010~A2 and other
active main belt asteroids suggest that a plausible catastrophic
disruption scenario consists of a small (\eg\ $100\meter$) diameter
asteroid disrupted into visibility, decaying slowly enough that
corresponding tracklets are reported to the MPC by multiple surveys,
after which the object is designated a `new' main belt
asteroid. Observations of such an object would not appear in the MPC's
ONS file as subsequent observations could be associated with the
initial orbit of the disrupted asteroid. The signature of a disruption
would be its sudden appearance (\ie\ the object would not be present
in precovery images where one would expect to find it if it were a
normal asteroid) followed by the secular decay of its measured
absolute magnitude and, of course, associated morphological
behavior. The discovery of such an object would improve our
understanding of brightness decay rates of objects undergoing
disruption or further restrict the calculated rate limit if no objects
are discovered.  \citet{Cikota2014} mined the MPC observation catalog
to search for low-level activity in the main belt and found four
candidate asteroids not previously known to exhibit activity. Their
success suggests that a search for sudden turn-on objects may turn up
several candidate disruptions.

The slow-decay scenario (\ie\ small $\tau$) with followup is a class
of disruption behavior that would not be detected using our technique
that specifically requires the event to appear in the MPC ONS file.
These disruptions would be visible for years and be `discovered' as
main belt asteroids and therefore be omitted from that file. Thus, if
this scenario is common it would lead us to under-estimate the
disruption rate. Such an event can not exist in the \PSone\ dataset
due to the small likelihood of the survey obtaining self-followup of
any object in our time window (\S\ref{ss.PanSTARRS1}).  The improving
efficacy of contemporary all-sky surveys in identifying unusual
photometric and morphological behavior in asteroid detections
(\eg\ (593)~Scheila, P/2012~F5~(Gibbs), P/2013~P5 and P/2013~R3) suggests that when such an
event occurs in the main belt, the event would be detected and
verified as a disruption and could be used in place of our candidate
bright ONS tracklet to measure $H_0$ (this would also require an
analysis of the survey's catastrophic disruption detection efficiency that allows for
discovery of this type of catastrophic disruption event).

\section{Conclusions}
\label{ss.Conclusion}

We have set limits on the rate of impact-generated catastrophic
disruptions in the main belt using the detection of the brightest
unknown object that had main belt-like motion (with $V=\vnought$)
during 453 days of sky surveying with \PSone.  That object must either
be a generic but unknown main belt asteroid with $H\sim16.5$ or an
unknown small asteroid undergoing a transient increase in brightness.
We can not determine what type of `activation' process might have
created the event but, since catastrophic disruptions are the least
likely of all of them, assuming that it is a catastrophic disruption provides a conservative
upper limit on the disruption rate.  Our catastrophic disruption rate
limit $H_0^{CL}(\Delta m,\tau)$ is stated in terms of the absolute
magnitude $H_0$ at which one main belt asteroid is catastrophically
disrupted per year and is provided as a function of the magnitude
increase due to the disruption ($\Delta m$) and the magnitude decay
rate ($\tau$) during the post-disruption time period as the remnant
dust cloud dissipates.  Our simplistic model of the photometric behavior of an
impact-generated catastrophic disruption suggests that $\Delta m \sim
+20$ and $0.01\magperday \la \tau \la 0.1\magperday$ which leads to
the conclusion that our 90\% confidence limit requires that $H_0^{CL} \ga
28.7$.

Our $H_0^{CL}$ limit is in strong disagreement with the expected value of
$H_{0,Bottke} \sim 23.3$ from contemporary main belt collisional evolution
models \citep[\eg][]{Bottke2005,OBrien2003}.  Their $H_{0}$ value requires $11.0\mags \la \Delta m \la
12.4\mags$ if they are to occur once per year. Even the uppermost value in this range is much smaller
than the $\Delta m=+20$ predicted by our simple dust model for the
brightness increase in the aftermath of an impact-generated
catastrophic disruption of a $100\meter$ diameter asteroid and from
modeling the impact-generated disruption scenarios for asteroid
P/2010~A2 {\citep{Jewitt2010}.  

The discrepancy between our measured $H_0^{CL}=28.7$ modeled using
impact-scenario brightness profiles and the expected
$H_{0,Bottke}=\bottkehnoughtfitted$, in combination with the recently observed rotation-induced
disruptions of main belt asteroids, suggests that the dominant cause of disruption for small asteroids and
is non-collisional, namely YORP spin-up disruption.  The discrepancy
can be reconciled by allowing spin-up disruptions, which are $400\times$ more
common than impact disruptions, to have brightness increases less than
those for impact-generated events but with similar decay rates.  We find then
that overall disruption rate (impact and rotational) is then
consistent with the predictions of \citet{Bottke2005}.
We can assume that a rotational breakup of a $D\simeq100$ m rubble pile asteroid occurs at a spin period of $P\simeq2.2$ hours \citep{Harris1996}. If this occurs then equatorial material will leave the asteroid at the rotation speed of $\sim2-3\mps$, similar to that assumed in our collisional disruption model. However it may be that such disruptions occur over long periods, with the asteroid gradually fragmenting via episodic mass-loss as appears to be the case for P/2013 P5. In this case the magnitude increase at any time within the photometric aperture will be far smaller than for an instantaneous disruption.

The steadily improving capability of asteroid surveys such as
\PSone\ and the Catalina Sky Survey are working towards the constant
monitoring of main belt asteroids in the night sky.  Their ability to
detect fainter, smaller asteroids will enable deeper completeness of
the main belt SFD, which in turn allows for easier and more frequent
detection of anomalous activity. These surveys are already detecting a
handful of unusual events per year, and upgrades to the
detection capability of both surveys could improve this number by a
factor of several by mid-2015, with additional improvement from 
surveys currently under construction.  The detection of an unambiguous
catastrophic disruption event is imminent.

The ATLAS survey \citep{Tonry2011}, dedicated to scanning the entire
sky nightly to search for `death-plunge' asteroids, will provide
additional monitoring of the bright end of the main belt.  It will be
capable of accurate photometry of sources as bright as $V=6$, a regime
where bright, rapidly-fading disruptions can not be monitored by
current surveys because their detectors saturate at much fainter
magnitudes.

Active main belt asteroids with unusual morphologies such as
P/2010~A2, P/2013~P5, and P/2013~R3 are guiding the way to more
sophisticated dust and fragment modeling of disruption
scenarios. These models, combined with a future steady stream of
disrupted objects and improved survey characterization, will lead to a
better understanding of main belt evolution based upon a wealth of
observational data complemented by more sophisticated dynamical
evolution and catastrophic disruption models.

\acknowledgements
\section*{Acknowledgments}

The \PSone\ Survey has been made possible through contributions of the
Institute for Astronomy, the University of Hawaii, the Pan-STARRS
Project Office, the Max-Planck Society and its participating
institutes, the Max Planck Institute for Astronomy, Heidelberg and the
Max Planck Institute for Extraterrestrial Physics, Garching, The Johns
Hopkins University, Durham University, the University of Edinburgh,
Queen's University Belfast, the Harvard-Smithsonian Center for
Astrophysics, and the Las Cumbres Observatory Global Telescope
Network, Incorporated, the National Central University of Taiwan, and
the National Aeronautics and Space Administration under Grant
No. NNX08AR22G issued through the Planetary Science Division of the
NASA Science Mission Directorate.

H.H.H. was supported by NASA through Hubble Fellowship Grant HF-51274.01 awarded by the Space Telescope Science Institute, which is operated by the Association of Universities for Research in Astronomy (AURA) for NASA, under contract NAS 5-26555.

Jan Kleyna was supported by NSF Award 1010059.

We thank David Jewitt and William Bottke, Jr. for their candid and insightful comments.

\clearpage


\section*{Appendix}
\label{sec.appendix}

Assume we have an asteroid of diameter $D_0$ that disrupts instantaneously into a `cloud' of particles of diameter $d$ that expands linearly in time with speed $2\,v$.  \ie\ the fastest particles are moving outwards at speed $v$ and the diameter of the cloud of particles at time $t$ is $D(t)=D_0+2vt$.  In this model the total number of particles in the cloud is
\begin{equation}
N = \Bigg({D_0 \over d}\Bigg)^3.
\end{equation}

Let $F_0$ represent the total light intensity flux from the original asteroid.  \ie\ from an asteroid with cross-sectional area $A_0 = \pi D_0^2 / 4$.  Thus, in an `aperture' of diameter $D_{aperture}>D_0$ the total flux is $F_0$ at time $t=0$.

The expanding dust cloud is optically thick while the cross-sectional area of all the particles ($A_{particles} = N a_{particle} = N \pi \, d^2 / 4$) is greater than the cross-sectional area of the cloud ($A_{cloud} = \pi D^2 / 4$).  Thus, the time until which the cloud is optically thick is given by
\begin{equation}
N d^2 \ge ( D_0 + 2 \, v \, t_{thick} )^2
\end{equation}
or
\begin{eqnarray}
t_{thick} &=&    {1 \over 2 \, v} \Bigg( \sqrt{D_0^3 \over d} - D_0 \Bigg) \\
          &\sim& {1 \over 2 \, v}        \sqrt{D_0^3 \over d}
\end{eqnarray}
since $D_0 \gg d$.

Our canonical disruption of a $100\meter$ diameter asteroid into micron-sized particles ($10^{-6}\meter$) in a cloud expanding at $1\meter/\second$ remains optically thick for about $t_{thick} \sim 6\days$ after which it is about $1,000\km$ in diameter.  Our nominal photometry aperture $\theta_{aperture}=3\arcsec$ corresponds to about $D_{aperture} = \theta \, \Delta \sim 3,300\km$ at a geocentric distance of $\Delta=1.5\au$ typical of main belt asteroids so the cloud begins to thin before it reaches the aperture diameter.  The time at which the cloud fills the aperture is given by
\begin{equation}
t_{aperture} = { \theta \, \Delta \over 2 \, v }
\end{equation}
and for most cases $t_{thick}<t_{aperture}$.  In our nominal situation $t_{aperture} \sim 1.65\times10^6\second \sim 19\days$

For $t \le t_{thick}$ the flux in the aperture relative to the initial flux increases like the cross-sectional area of the dust cloud:
\begin{equation}
f(t) = {F(t) \over F_0} 
     = \Biggl[ {  D(t)   \over D_0 } \Biggr]^2 
     = \Biggl( { D_0+2vt \over D_0 } \Biggr)^2
     = \Biggl( 1 + { 2vt \over D_0 } \Biggr)^2.
\end{equation}
We use a small $f$ to represent the flux relative to the initial flux.  In our nominal case $2vt \gg D_0$ so that
\begin{equation}
f(t) \sim \Biggl( { 2vt \over D_0 } \Biggr)^2.
\end{equation}

For $t_{thick} < t \le t_{aperture}$ the flux in the aperture relative to the initial flux remains constant because, even though the cloud is optically thin, all the light is still contained in the aperture: 
\begin{equation}
f(t) = N { \pi d^2 / 4 \over \pi D_0^2 / 4 } 
     = \Bigg({D_0 \over d}\Bigg)^3 { d^2 \over D_0^2 }
     = {D_0 \over d}.
\end{equation}
Our nominal case yields a $10^8$ increase in brightness of the object corresponding to $\Delta V=20$!

Once $t > t_{aperture}$ the dust is optically thin and the flux in the aperture is only contributed by the dust {\it within} the aperture.  The `aperture' is a `cylindrical hole' through a sphere where the cylinder has a diameter $D_{aperture}$ with spherically shaped ends (see fig.~\ref{fig.SphericalCap}).   We assume that the dust is homogeneously distributed in the cloud so that the number of particles in the aperture is $n(t) = N \, V_{aperture}(t) / V_{cloud}(t)$ where $V_{aperture}(t)$ and $V_{cloud}(t)={4\over3}\pi [D(t)/2]^3$ are the volumes of the aperture and the entire cloud respectively.  Letting $r_a=D_{aperture}/2$ and $R(t)=D(t)/2$ but omitting the time dependence for clarity:
\begin{eqnarray}
V_{aperture} 
  &=& V_{cylinder}(r_a,R) + 2 \, V_{cap}(r_a,R) \\ 
  &=& 2 \pi  r^2 \sqrt{R^2-r^2}+\frac{1}{3} \pi  \left(R-\sqrt{R^2-r^2}\right) \left(\left(R-\sqrt{R^2-r^2}\right)^2+3 r^2\right)
\end{eqnarray}

For $t > t_{aperture}$ the flux in the aperture relative to the initial flux is simply the ratio of the cross-sectional area of the particles in the aperture to the cross-sectional area of the original asteroid.  Computing the ratio and substituting $R(t) \simeq vt$:
\begin{eqnarray}
f(t) &=& { n(t) \, a_{particle} \over \pi (D_0/2)^2 }\\
     &=& { n(t) \Bigg( { d \over D_0 } \Bigg)^2 }\\
     &=& { N \, { V_{aperture}(t) \over V_{cloud}(t) } 
          \Bigg( { d \over D_0 } \Bigg)^2 }\\
     &=& { { D_0 \over d } { V_{aperture}(t) \over V_{cloud}(t) } }\\
     &=&   { D_0 \over d } \Bigg[ 1-\frac{\left(t^2 v^2-r^2\right)^{3/2}}{t^3 v^3} \Bigg].
\end{eqnarray}

To summarize the time behavior of the flux relative to the original flux:
\begin{equation}
f(t) = 
\left\{
\begin{array}{l r l}
{(D_0+2vt)^2 \over D_0^2 } \sim { 4 v^2 \over D_0^2 } \, t^2  & 0<&t<t_{thick} \\
{D_0 \over d}                                 & t_{thick} < &t \le t_{aperture} \\
{ { D_0 \over d } \Bigg[ 1-\frac{\left(t^2 v^2-r_a^2\right)^{3/2}}{t^3 v^3} \Bigg]} & & t > t_{aperture} \\
\end{array}
\right.
\label{eq.flux.vs.time}
\end{equation}

The change in magnitude in the photometry aperture is $\Delta m(t) = - 2.5 \, \log_{10} f(t)$:
\begin{equation}
\Delta m(t) = 
\left\{
\begin{array}{l r l}
   - 2.5 \, \log_{10} \Big[{ 4 v^2 \over D_0^2 } \, t^2                                  \Big] &           0<&t<t_{thick} \\
   - 2.5 \, \log_{10} \Big[{D_0 \over d}                                                 \Big] & t_{thick} < &t \le t_{aperture} \\
   - 2.5 \, \log_{10} \Big[{ { D_0 \over d } \Big[ 1-\frac{\left(t^2 v^2-r_a^2\right)^{3/2}}{t^3 v^3} \Big]} \Big] &             &t > t_{aperture} \\
\end{array}
\right.
\label{eq.deltaM}
\end{equation}

Thus, in our model the magnitude decreases (gets brighter) like $\log \, t$ for $t<t_{thick}$, remains constant for $t_{thick} < t < t_{aperture}$ and finally increases (gets fainter) like $\log \, t$ for $t\gg t_{aperture}$.

The rate of change in magnitude in the photometry aperture is 
\begin{equation}
{\dif \Delta m(t) \over \dif t} = {2.5 \over \ln(10) f(t)} {\dif f(t) \over \dif t}
\end{equation}
so that
\begin{equation}
{\dif \Delta m(t) \over \dif t} = 
\left\{
\begin{array}{l r l}
   - { 5 \over \ln{10} } \, {1 \over t} \sim - { 2 \over t } &           0<&t<t_{thick} \\
   0                                                            & t_{thick} < &t \le t_{aperture} \\
   -\frac{3 {2.5 \over \ln(10)} r^2 \sqrt{t^2 v^2-r^2}}{t \left(t^2 v^2 \sqrt{t^2 v^2-r^2}-r^2 \sqrt{t^2 v^2-r^2}-t^3 v^3\right)} &             &t > t_{aperture} \\
   + { 2.5 \over \ln{10} } \, {2 \over t} \sim + { 2 \over t } &             &t \gg t_{aperture} \\
\end{array}
\right.
\label{eq.magnitude.vs.time}
\end{equation}
Thus, $200\days$ after the catastrophic disruption the apparent magnitude in the photometry aperture is increasing at a rate of $\sim 0.01\magperday$. Figs. \ref{fig.mag} and \ref{fig.dmag} show the behavior of the change in brightness and rate of change in brightness using our example disruption and a range of particle expansion velocities. We find that for our 453-day survey window and a reasonable range of expansion velocities, the magnitude decay rate spends most of its time between $0.01$ and $0.1\magperday$.

%



\clearpage
\begin{deluxetable}{lc}
\tabletypesize{\small}
\tablecaption{\tcd\ Observational Parameters}
\tablehead{
\colhead{Parameter} & 
\colhead{Value}
} 
\startdata
(R.A., declination) (J2000) & ($21^{h}22^{m}39^{s}, \, -19\arcdeg48\arcmin12\arcsec$)  \\
Ecliptic longitude \& latitude w.r.t. Opposition & ($0.5\arcdeg, \, -2.0\arcdeg$) \\
Rate of motion & 0.24 deg~day$^{-1}$\\
\digest\ scores\tablenotemark{1} & MB1 $=42$, MB2 $=39$, MB3 $=5$ \\
Great-circle residual & $0.02\arcsec$ \\
Absolute $V$-magnitude\tablenotemark{2} & 16.5 \\
Heliocentric distance\tablenotemark{2} & $1.95\bigau$ \\
\enddata
\tablenotetext{1}{See \S\ref{ss.Search}.}

\tablenotetext{2}{From the maximum likelihood orbit computed by
  OpenOrb \citep{Granvik2009}. The orbit and digest score were
  computed using only the first two detections of the three-detection
  tracklet because the third detection was contaminated by an
  electronic artifact.}

\label{t.params}
\end{deluxetable}

\clearpage
\begin{deluxetable}{lcccccc}
\tabletypesize{\small}
\tablecaption{Recent `activated' Solar System objects}
\tablehead{
\colhead{Object} & 
\colhead{Type} & 
\colhead{Diameter} &
\colhead{Period of} &
\colhead{Nature of} &
\colhead{$\Delta m$} &
\colhead{$\tau$} \\
\colhead{} & 
\colhead{} & 
\colhead{(km)} &
\colhead{activity} &
\colhead{activity} &
\colhead{(mag)} &
\colhead{(m~day$^{-1}$)}
} 
\startdata 
P/2010~A2\tablenotemark{1} & MBC\tablenotemark{\ast} & 0.12 & 2009-2011 & Collision & $<19$ & 0.055 \\
P/2010~A2\tablenotemark{1} & MBC\tablenotemark{\ast} & 0.12 & 2009-2011 & Rotational breakup & 3.9 & 0.014 \\
(596) Scheila\tablenotemark{2} & MBA\tablenotemark{\dagger} & 113 & 2010 & Crater formation & 1 & 0.03 \\
17P/Holmes\tablenotemark{3} & Comet & N/A & 2007-2008 & Sublimation(?) & 15 & 0.04 \\
P/2012 F5 (Gibbs)\tablenotemark{4} & MBC\tablenotemark{\ast} & 2.0 & 2011-2012 & Crater formation & 2.0 & 0.002 \\
P/2013 P5\tablenotemark{5} & MBC\tablenotemark{\ast} & 0.24 & Apr.-Sep. 2013 & Episodic rotational & N/A & N/A \\
P/2013 R3\tablenotemark{6} & MBC\tablenotemark{\ast} & Multiple\tablenotemark{7} & Sep.-Dec. 2013 & Rotational breakup & 4.0 & 0.03 \\
\enddata
\label{t.activeobjects}
\tablenotetext{\ast}{Main belt comet \citep{Hsieh2006}}
\tablenotetext{\dagger}{Main belt asteroid}
\tablenotetext{1}{Computed using the upper limit of $\Delta m$ from \citet{Jewitt2012} over a 9-month interval.}
\tablenotetext{2}{\citet{Bodewits2011}}
\tablenotetext{3}{\citet{Stevenson2012a}}


\tablenotetext{4}{\citet{Novakovic2014}, \citet{Stevenson2012b}}
\tablenotetext{5}{\citet{Jewitt2013b}. The effect of multiple, possibly layered, episodes of regolith release for P/2013~P5 preclude a meaningful notion for brightness increase $\Delta m$ or decay $\tau$.}
\tablenotetext{6}{\citet{Jewitt2014}}
\tablenotetext{7}{P/2013~R3 was discovered already split into at least 10 fragments, with the largest having an effective diameter $\la$ 400~m. We crudely estimated $\Delta m$ using brightness decay measured by \citet{Jewitt2014} extrapolated to an average disruption date between February and September 2013.}
\end{deluxetable}

\clearpage
\begin{deluxetable}{lccccc}
\tabletypesize{\small}

\tablecaption{Estimated number of catastrophic disruptions per year
  brighter than $V = \vnought$ that will be detected by ongoing and
  anticipated asteroid surveys.}

\tablehead{
\colhead{Survey} & 
\colhead{Dates of Operation} & 
\colhead{Survey Efficiency\tablenotemark{\star}} &
\colhead{Mag Range} &
\colhead{Disruptions/Year}
} 
\startdata
\PSone\             & 2010-2013  & 0.4       & 13-22   & 0.8 \\
\PSone\             & 2014       & 0.9       & 13-22   & 2 \\
\PSone+2             & 2015-2016 & 1        & 13-22   & 2 \\
Catalina Sky Survey & 2014-      & 1        & 13-22   & 2 \\
ATLAS               & 2016-2018  & 2        & 6-20    & 10 \\
SST\tablenotemark{\dagger} & 2015(?)-   & 2 & 13-22   & 4 \\
LSST                & 2021(?)-      & 2        & 18-24.5   & 0\tablenotemark{\ddag} \\
\enddata
\label{t.ndetectedbysurvey}
\tablenotetext{\dagger}{Estimated from \citet{Shah2013}.}
\tablenotetext{\ddag}{The LSST will have a saturation limit of $V\sim18$ \citep{Kantor2013} ensuring that nearly zero $V<\vnought$ disruptions will be detected.}
\tablenotetext{\star}{Our measure of a survey's ability to detect catastrophic disruptions near opposition over a single lunation. Efficiency $> 1$ indicates repeat visits to the same region over successive lunations.
}
\end{deluxetable}


\clearpage
\begin{sidewaysfigure}[btp]
\centering
\includegraphics[scale=0.7]{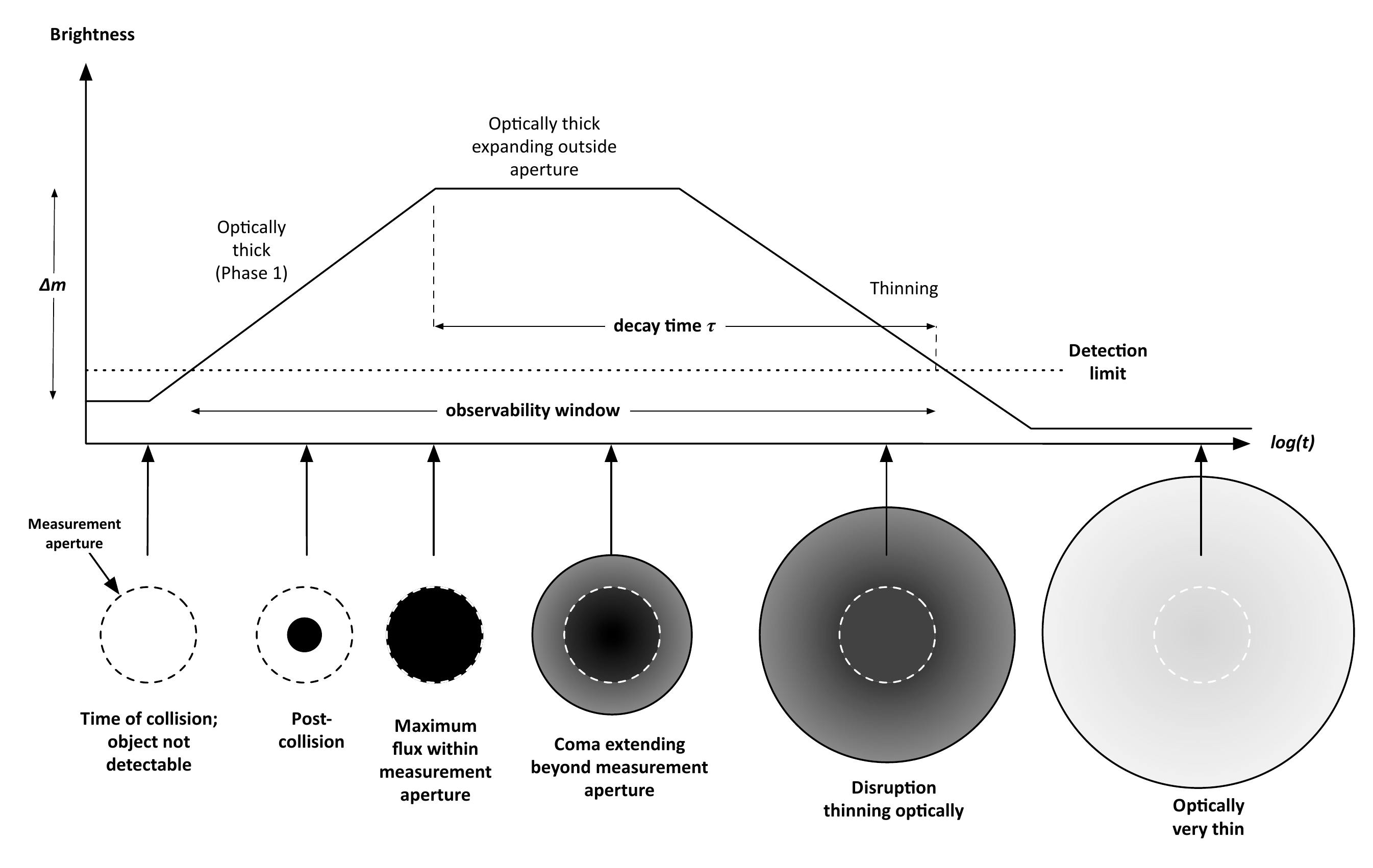}
\caption{Our model (bottom) of a catastrophic disruption and its
  associated photometric behavior in a small fixed aperture (top). For this
  example the disruption plateau occurs when the dust cloud is optically thick while expanding beyond the
  measurement aperture. A plateau can also occur if the dust cloud is thinning but
  is completely enclosed by our measurement aperture. Out
  model is discussed in detail in \S\ref{ss.DisruptionModel}.}
\label{fig.modeldisruption}
\end{sidewaysfigure}

\clearpage
\begin{figure}[btp]
\centering
\includegraphics[scale=0.6]{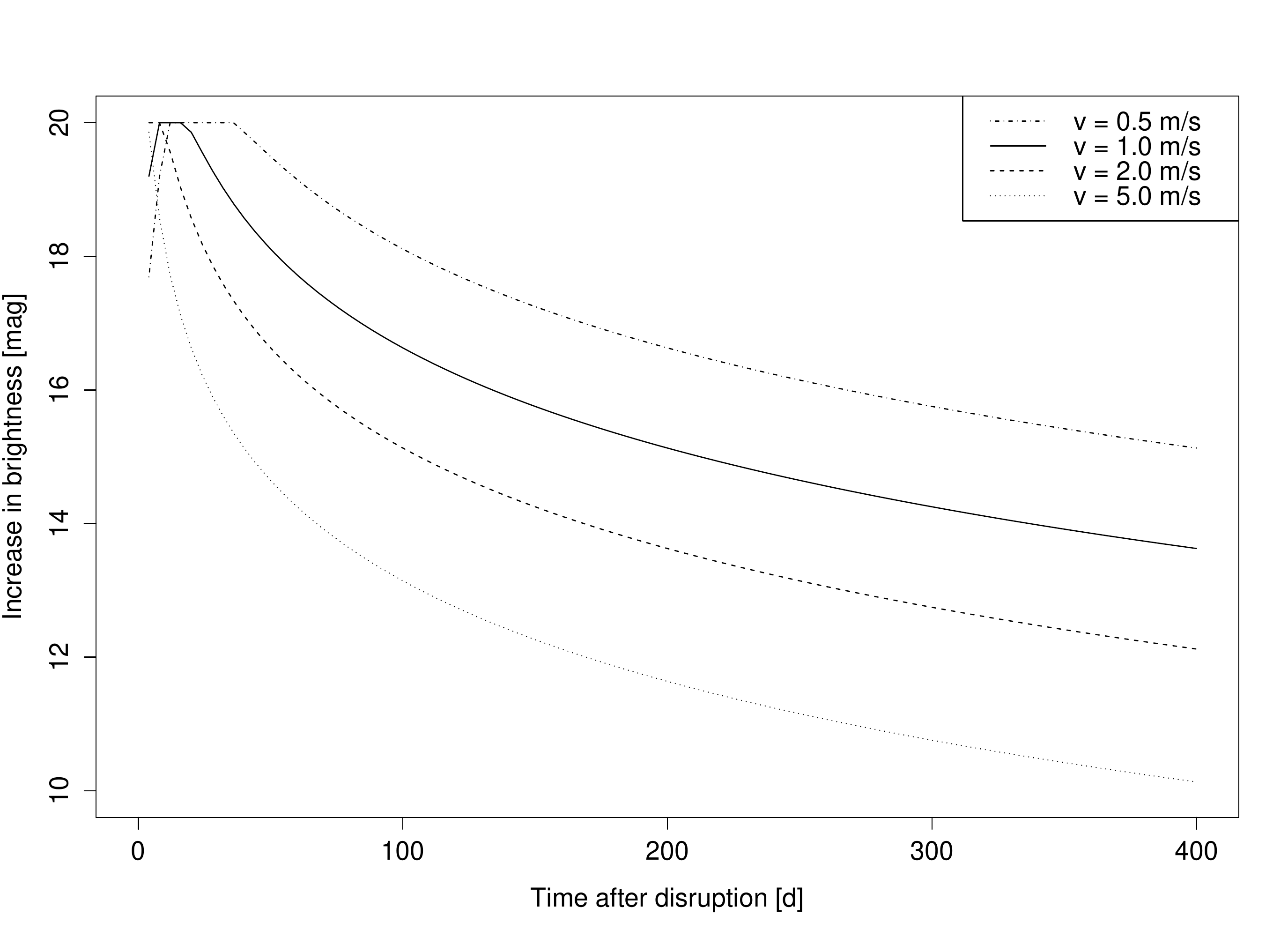}
\caption{The increase in brightness of a model catastrophic disruption for a range
of dust dispersion velocities. Our nominal case of a 100 m parent body converted to $1\um$ particles results in a $+20$ increase in magnitude. In all cases, a sharp increase in brightness occurs, followed by a brief plateau, then a decay in brightness. While the magnitude profiles are not linear, a sawtooth profile (instant rise in brightness followed by linear decay) captures the essential behavior for simulation purposes.}
\label{fig.mag}
\end{figure}

\clearpage
\begin{figure}[btp]
\centering
\includegraphics[scale=0.6]{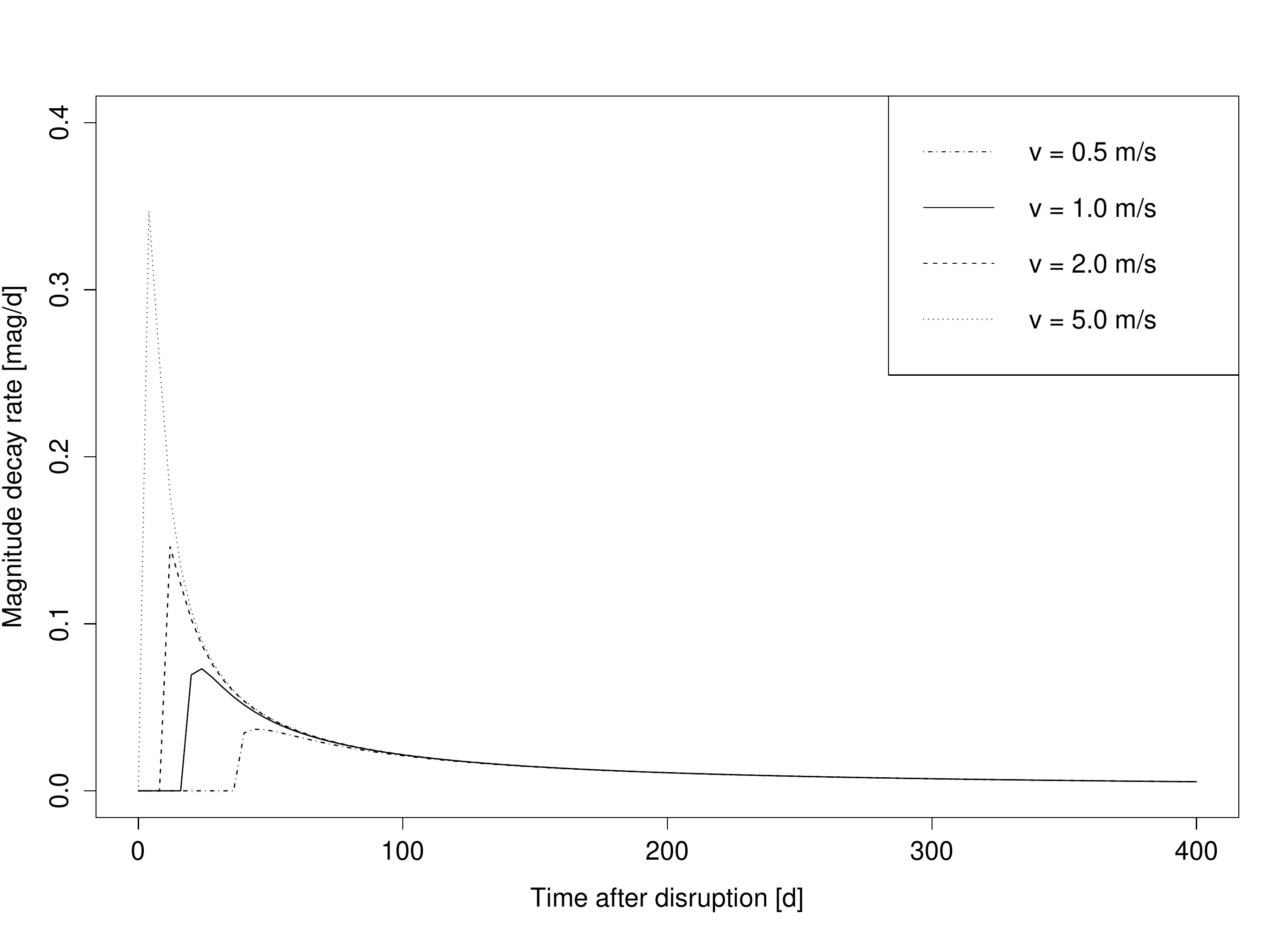}
\caption{Time derivatives for the brightness profiles of fig. \ref{fig.mag}. Except for rapid initial decay of large (\eg\ $5\mps$) dust dispersion velocities,
all our model disruptions exhibit brightness decays from $\sim0.1$ to $0.01\magperday$.}
\label{fig.dmag}
\end{figure}

\clearpage
\begin{figure}[btp]
\centering
\includegraphics[scale=0.5]{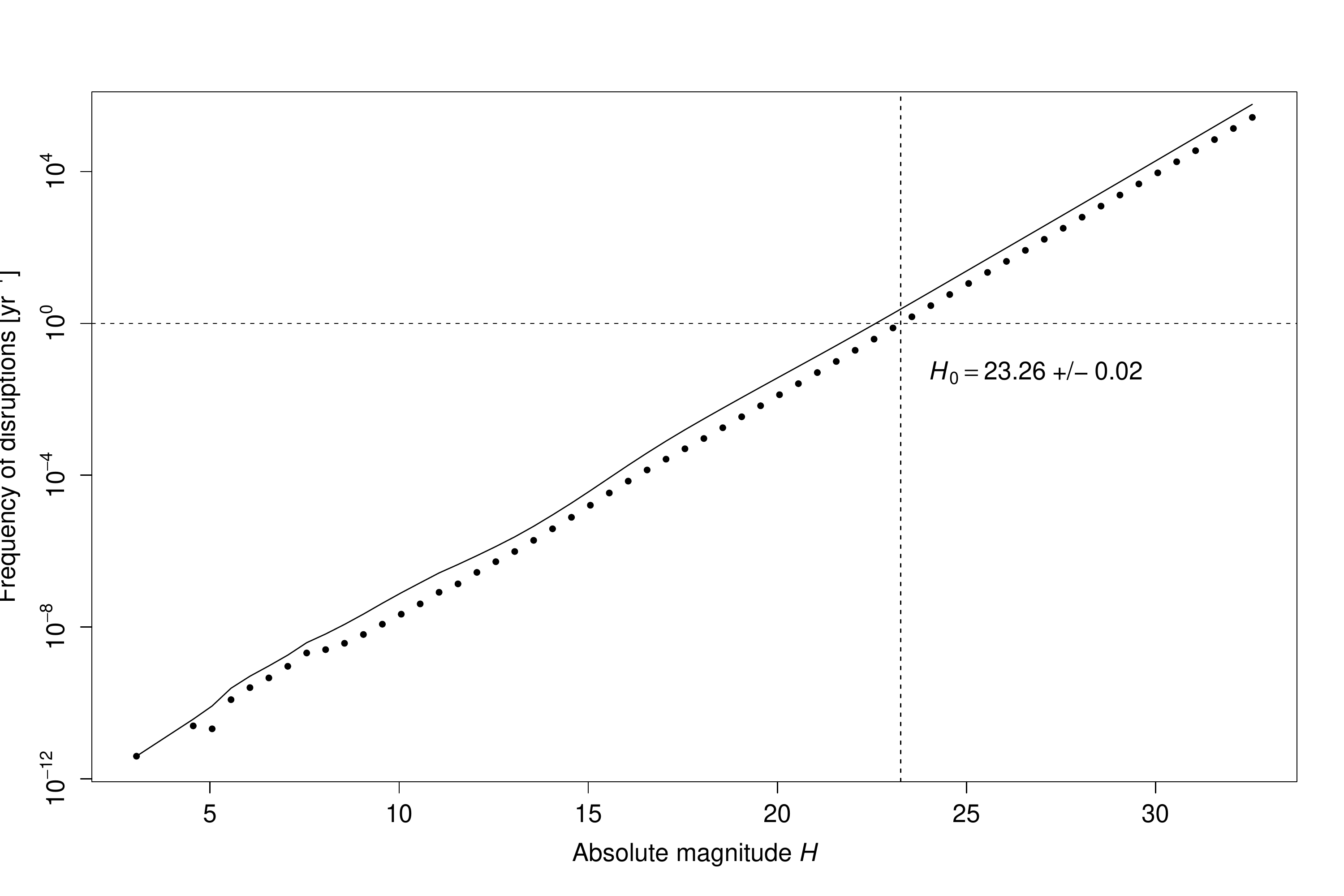}
\caption{Catastrophic main belt asteroid disruption event frequency as
  a function of absolute magnitude $H$ from \citet{Bottke2005}. The
  black dots represent the interval in each absolute magnitude
  bin. The solid line represents the cumulative disruption interval
  for asteroids larger than a given absolute magnitude.  Our fit to
  the model yields a power-law slope of $\beta=0.57\pm0.002$ and
  suggests that one $\sim100\meter$ diameter main belt object is
  catastrophically disrupted each year corresponding to
  $H_0=\bottkehnoughtfitted\pm0.02$.}
\label{fig.bottke}
\end{figure}

\clearpage
\begin{figure}[btp]
\centering
\includegraphics[scale=0.6]{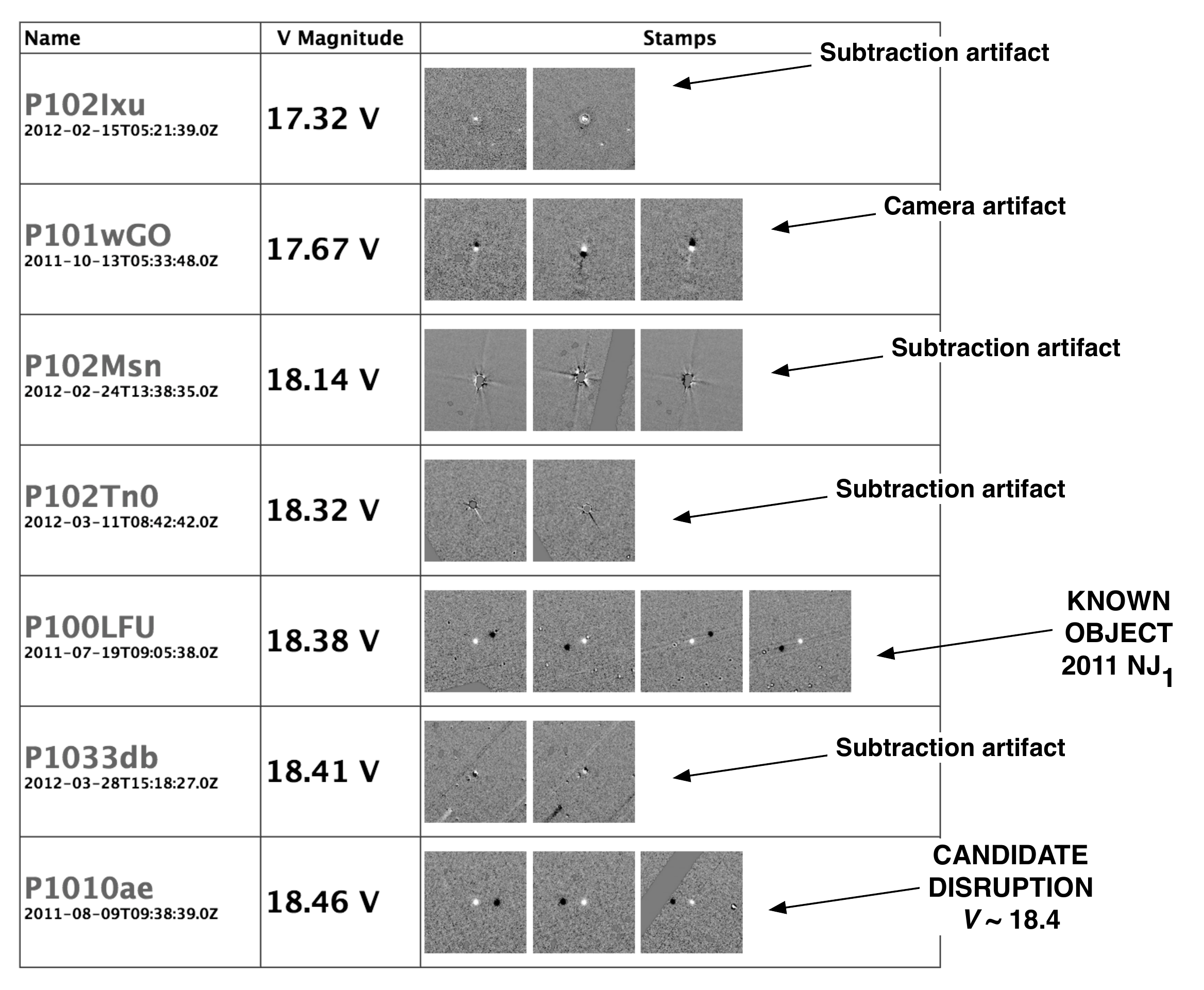}
\caption{200 $\times$ 200 pixel `postage stamps' of
  pairwise-subtracted images for \PSone\ tracklets in the MPC's `one
  night stand' file in order of increasing magnitude.  The brightest
  unknown real object, \tcd, has apparent magnitude
  $V\sim\vnought$. The solid grey areas represent detector gaps or
  masked pixels.}
\label{fig.onslist}
\end{figure}

\clearpage
\begin{figure}[btp]
\centering
\includegraphics[width=0.8\textwidth]{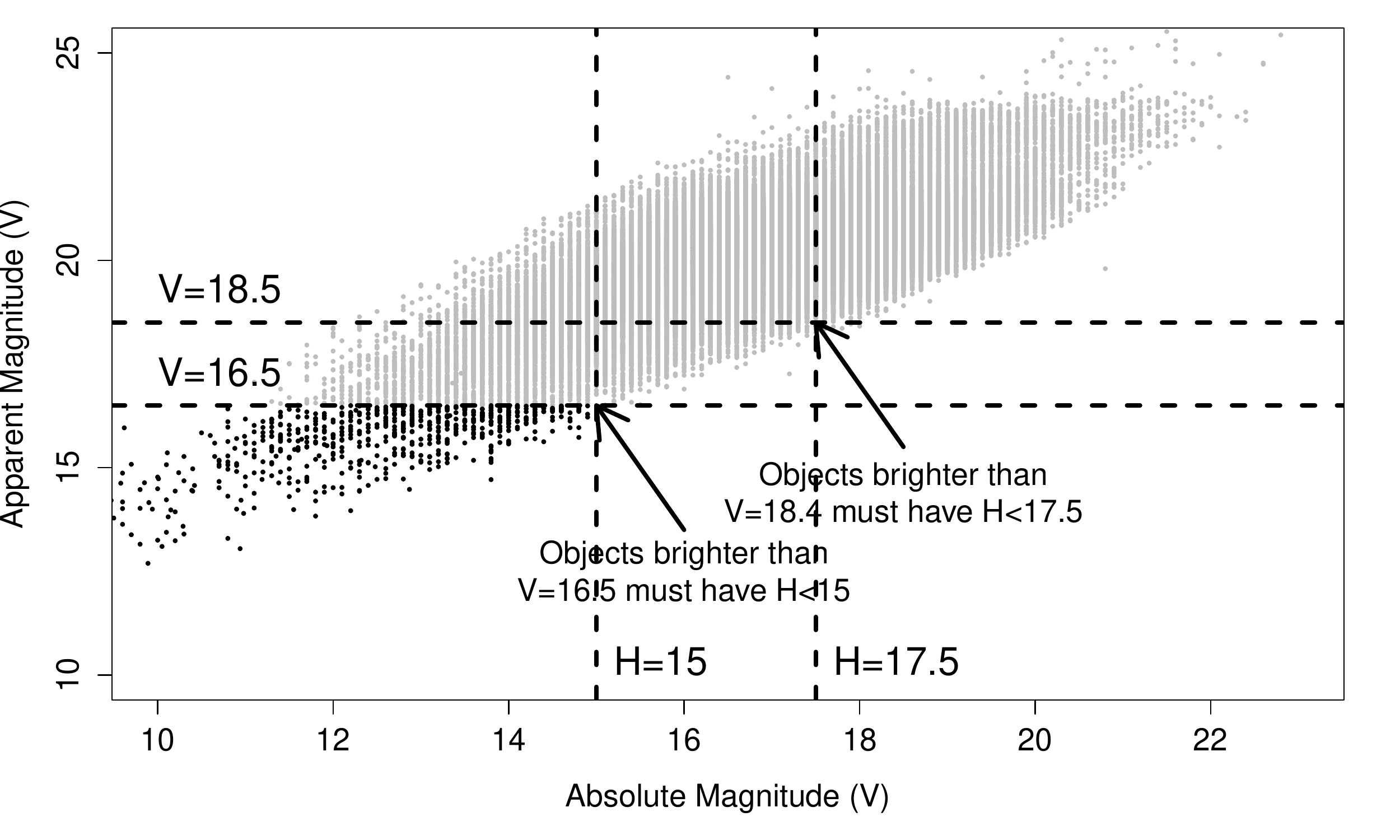}
\includegraphics[width=0.8\textwidth]{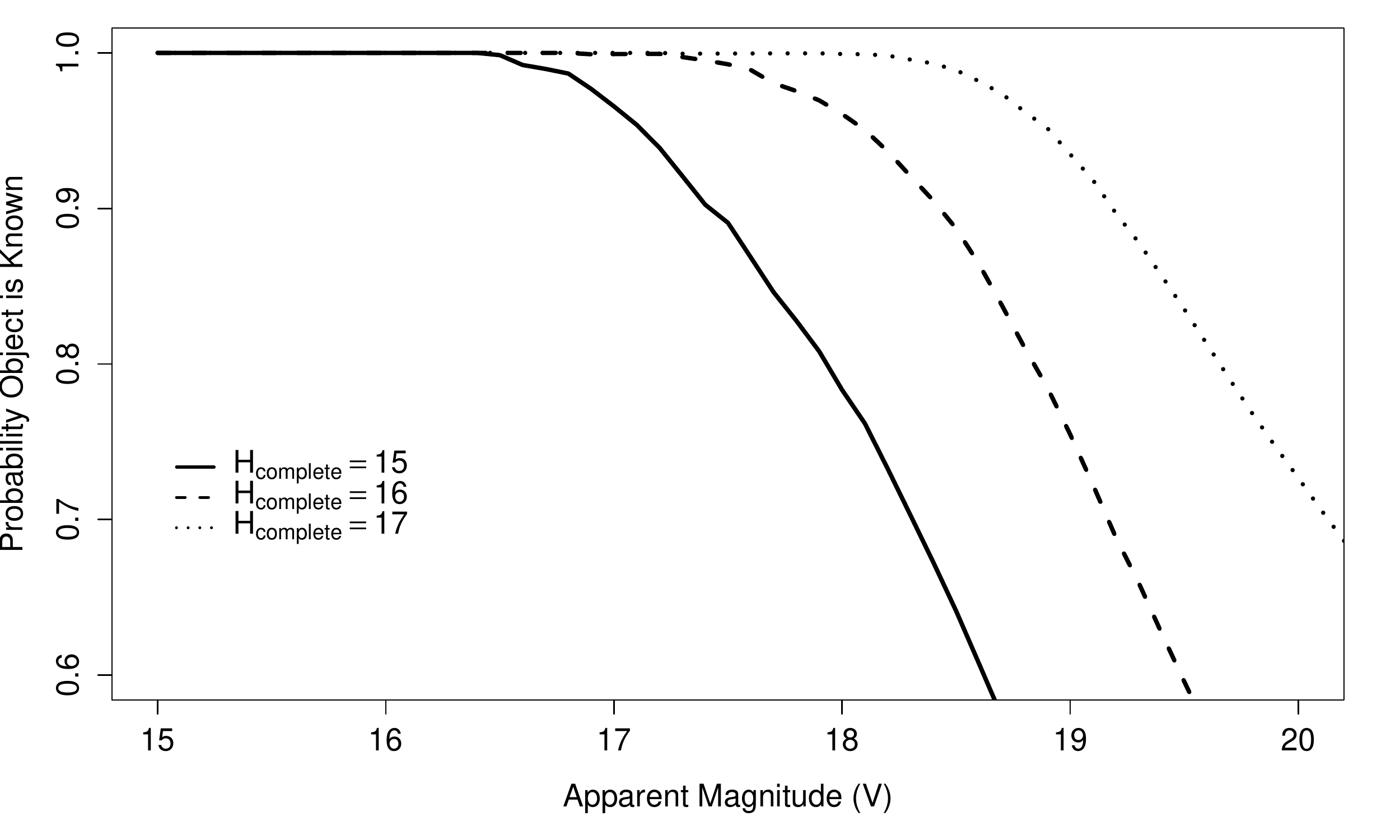}
\caption{(Top) Apparent $V$ magnitude vs. absolute $V$ magnitude of
  all known main belt asteroids within 30$^\circ$ of opposition on 1
  December 2013. The horizontal $V=16.5$ line corresponds to the
  faintest magnitude for the sample if the belt is complete to
  $H=15$. (Bottom) The probability a tracklet observed near opposition
  with main belt motion and a given apparent magnitude has
  $H<H_{complete}$ for $H_{complete}=15$, 16, and 17. \PSone\ has
  submitted no unknown tracklets brighter than $V=\vnought$.  }
\label{fig.hcompleteness}
\end{figure}

\clearpage
\begin{figure}[btp]
\centering
\includegraphics[scale=0.8]{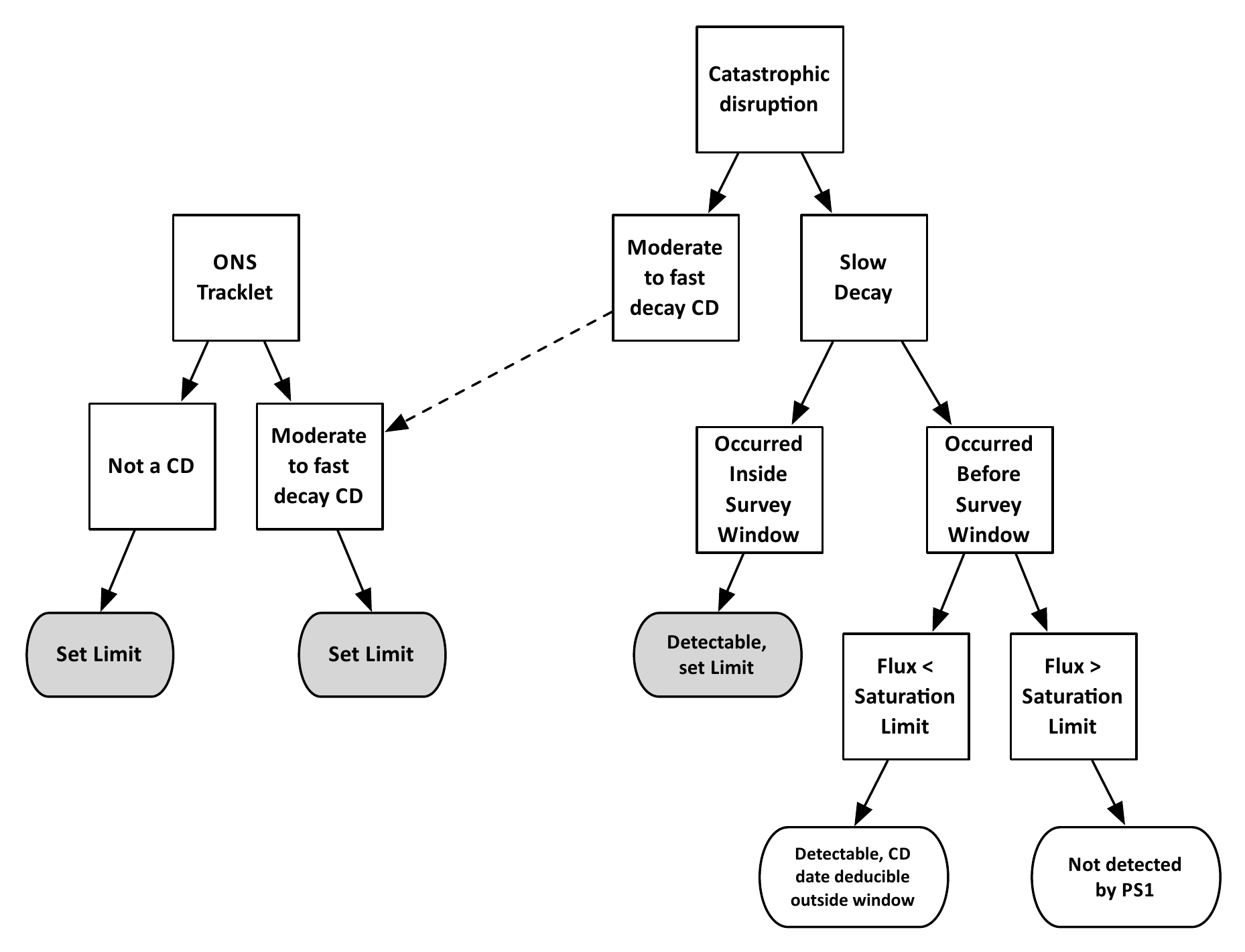}
\caption{The outcome tree that determines the route by which a catastrophic disruption can be detected by the \PSone\ system and determine a limit.  Our one-night-stand (ONS) search is limited to moderate- or fast-decaying catastrophic disruptions. We reason that slower-decaying disruptions in opposition brighter than our $V = \vnought$ brightest candidate limit would have bypassed the ONS process and have been detectable by morphology, as happened with P/2013~P5 and P/2013~R3, which were $\sim2$ magnitudes fainter at discovery.}
\label{fig.onsdecisiontree}
\end{figure}

%

\clearpage
\begin{figure}[btp]
\centering
\includegraphics[scale=0.6]{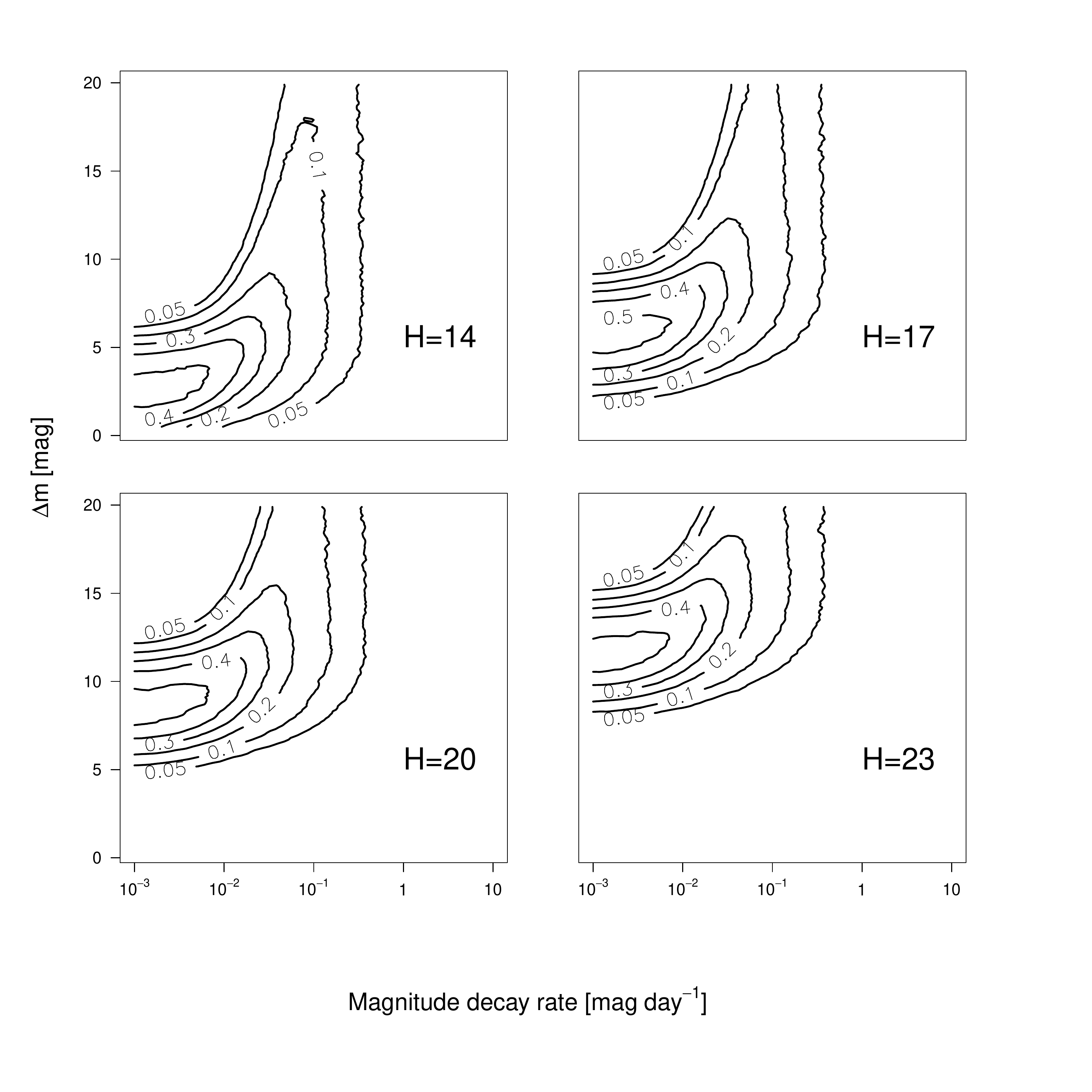}
\caption{Detection efficiency contours at $H=$ 14, 17, 20 and 23 for
  catastrophic disruptions as a function of the brightness increase
  $\Delta m$ and magnitude decay rate $\tau$ for \PSone\ pointings
  over the period 2011-02-21 through 2012-05-19.}
\label{fig.hgroup}
\end{figure}

\clearpage
\begin{figure}[btp]
\centering
\includegraphics[scale=0.60]{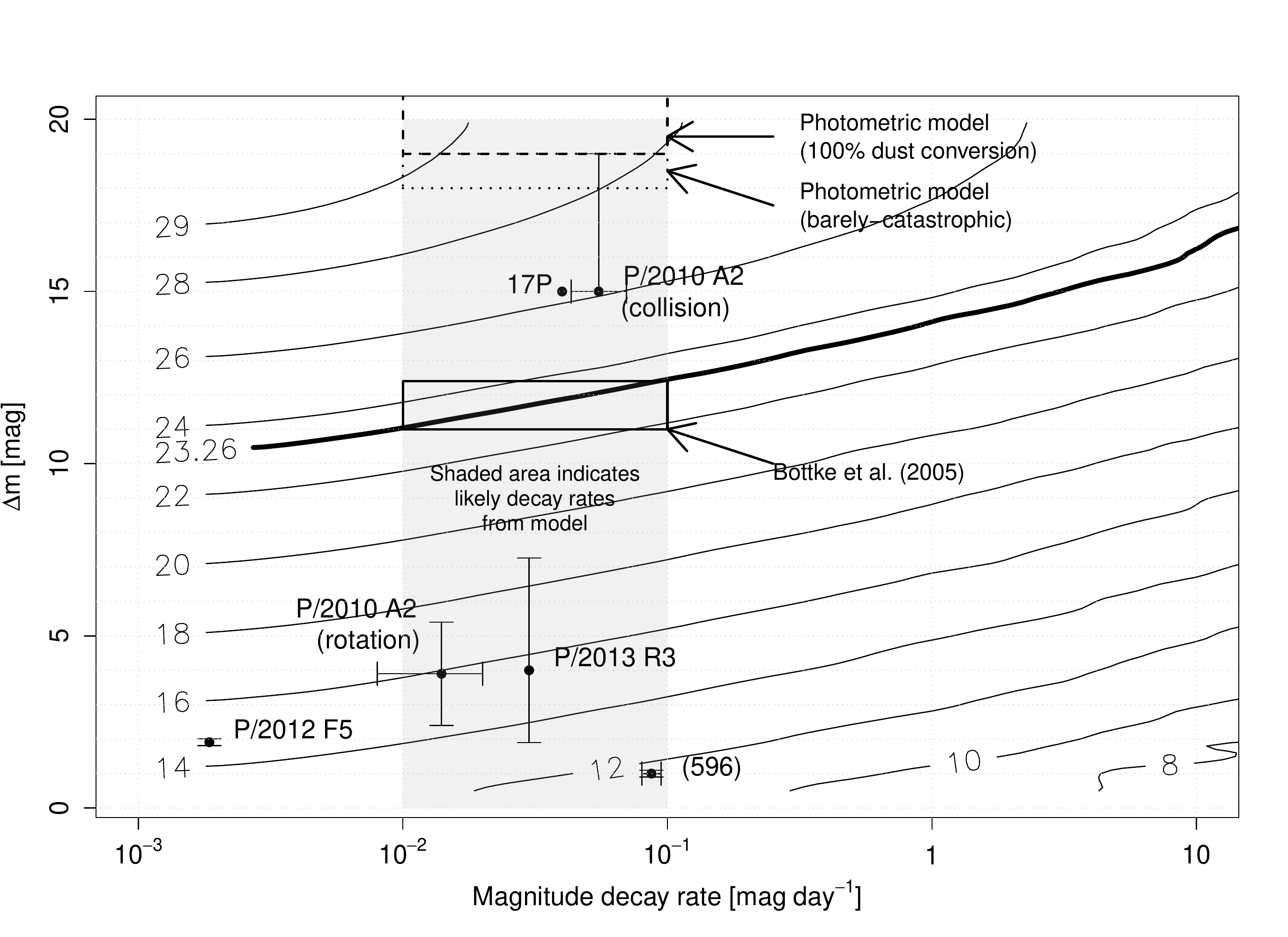}
\caption{90\% confidence limit contours for the absolute magnitude
  $H_{0}$ at which one catastrophic main belt asteroid disruption
  occurs per year.  The grey
  shaded area indicates the range of magnitude decay rates ($\tau$) we
  are likely to observe based on our photometric model. 
  The thick solid line corresponds to $H_0 = H_{0,Bottke} = \bottkehnoughtfitted$ \citep{Bottke2005}.
  The union of this line and our decay limits of $10^{-2}$ and $10^{-1}$ form a rectangle
  that bounds the brightness increase $\Delta m$ and decay rate $\tau$ for catastrophic
  disruptions if $H_0=\bottkehnoughtfitted$ is assumed. The dashed rectangle indicates
  brightness and decay behavior suggested by our simple collisional disruption model.  The discrepancy
  between these two regions suggests that either that 
  $H_{0,Bottke}$ is incorrect, or that bodies of absolute magnitude $H_{0,Bottke}=\bottkehnoughtfitted$ (100~m) are disrupting in a way
  that produces much more modest magnitude increases, \eg\ +11 to +13 magnitudes, than predicted by our simple model.
}
\label{fig.h0}
\end{figure}

\clearpage
\begin{figure}[btp]
\centering
\includegraphics[scale=.4]{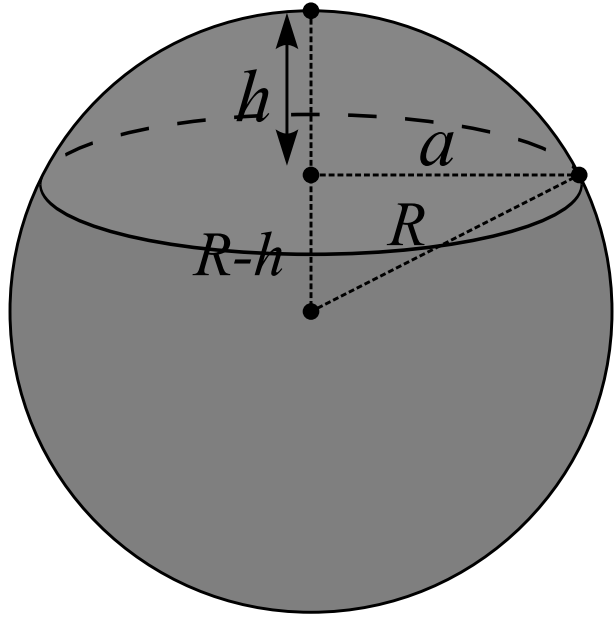}
\caption{Schematic of spherical cap figure to illustrate visible dust cross-section of catastrophic disruption \citep{SphericalCap}. The volume of the cap is 
$ V_{cap} = {\pi h \over 6} ( 3 a^2 + h^2 ) $. The total contribution of dust seen by a fixed measurement aperture consists of the particles contained in two end caps and the cylinder connecting them.
}
\label{fig.SphericalCap}
\end{figure}

\bibliography{references}

\begin{thebibliography}{}

\bibitem[{Agarwal} {\em et~al.}(2013){Agarwal}, {Jewitt}, and
  {Weaver}]{Agarwal2013}
{Agarwal}, J., D.~{Jewitt},\ and H.~{Weaver} 2013.
\newblock {Dynamics of Large Fragments in the Tail of Active Asteroid P/2010
  A2}.
\newblock {\em \apj\/}~{\bf 769}, 46.

\bibitem[{Bodewits} {\em et~al.}(2011){Bodewits}, {Kelley}, {Li}, {Landsman},
  {Besse}, and {A'Hearn}]{Bodewits2011}
{Bodewits}, D., M.~S. {Kelley}, J.-Y. {Li}, W.~B. {Landsman}, S.~{Besse},\ and
  M.~F. {A'Hearn} 2011.
\newblock {Collisional Excavation of Asteroid (596) Scheila}.
\newblock {\em \apjl\/}~{\bf 733}, L3.

\bibitem[{Bottke} {\em et~al.}(2005){Bottke}, {Durda}, {Nesvorn{\'y}},
  {Jedicke}, {Morbidelli}, {Vokrouhlick{\'y}}, and {Levison}]{Bottke2005}
{Bottke}, W.~F., D.~D. {Durda}, D.~{Nesvorn{\'y}}, R.~{Jedicke},
  A.~{Morbidelli}, D.~{Vokrouhlick{\'y}},\ and H.~F. {Levison} 2005.
\newblock {Linking the collisional history of the main asteroid belt to its
  dynamical excitation and depletion}.
\newblock {\em \icarus\/}~{\bf 179}, 63--94.

\bibitem[Bottke {\em et~al.}(2006)Bottke, Vokrouhlický, Rubincam, and
  {Nesvorn{\'y}}]{Bottke2006}
Bottke, W.~F., D.~Vokrouhlický, D.~P. Rubincam,\ and D.~{Nesvorn{\'y}} 2006.
\newblock The yarkovsky and yorp effects: Implications for asteroid dynamics.
\newblock {\em Annual Review of Earth and Planetary Sciences\/}~{\em 34\/}(1),
  157--191.

\bibitem[{Chapman} {\em et~al.}(2002){Chapman}, {Merline}, {Thomas}, {Joseph},
  {Cheng}, and {Izenberg}]{Chapman2002}
{Chapman}, C.~R., W.~J. {Merline}, P.~C. {Thomas}, J.~{Joseph}, A.~F. {Cheng},\
  and N.~{Izenberg} 2002.
\newblock {Impact History of Eros: Craters and Boulders}.
\newblock {\em \icarus\/}~{\bf 155}, 104--118.

\bibitem[{Cikota} {\em et~al.}(2014){Cikota}, {Ortiz}, {Cikota}, {Morales}, and
  {Tancredi}]{Cikota2014}
{Cikota}, S., J.~L. {Ortiz}, A.~{Cikota}, N.~{Morales},\ and G.~{Tancredi}
  2014.
\newblock {A photometric search for active Main Belt asteroids}.
\newblock {\em ArXiv e-prints\/}.

\bibitem[{Davis} {\em et~al.}(1989){Davis}, {Weidenschilling}, {Farinella},
  {Paolicchi}, and {Binzel}]{Davis1989}
{Davis}, D.~R., S.~J. {Weidenschilling}, P.~{Farinella}, P.~{Paolicchi},\ and
  R.~P. {Binzel} 1989.
\newblock {Asteroid collisional history - Effects on sizes and spins}.
\newblock In R.~P. {Binzel}, T.~{Gehrels}, and M.~S. {Matthews} (Eds.), {\em
  Asteroids II}, pp.\  805--826.

\bibitem[{Denneau} {\em et~al.}(2013){Denneau}, {Jedicke}, {Grav}, {Granvik},
  {Kubica}, {Milani}, {Vere{\v s}}, {Wainscoat}, {Chang}, {Pierfederici},
  {Kaiser}, {Chambers}, {Heasley}, {Magnier}, {Price}, {Myers}, {Kleyna},
  {Hsieh}, {Farnocchia}, {Waters}, {Sweeney}, {Green}, {Bolin}, {Burgett},
  {Morgan}, {Tonry}, {Hodapp}, {Chastel}, {Chesley}, {Fitzsimmons}, {Holman},
  {Spahr}, {Tholen}, {Williams}, {Abe}, {Armstrong}, {Bressi}, {Holmes},
  {Lister}, {McMillan}, {Micheli}, {Ryan}, {Ryan}, and {Scotti}]{Denneau2013}
{Denneau}, L., R.~{Jedicke}, T.~{Grav}, M.~{Granvik}, J.~{Kubica}, A.~{Milani},
  P.~{Vere{\v s}}, R.~{Wainscoat}, D.~{Chang}, F.~{Pierfederici}, N.~{Kaiser},
  K.~C. {Chambers}, J.~N. {Heasley}, E.~A. {Magnier}, P.~A. {Price},
  J.~{Myers}, J.~{Kleyna}, H.~{Hsieh}, D.~{Farnocchia}, C.~{Waters}, W.~H.
  {Sweeney}, D.~{Green}, B.~{Bolin}, W.~S. {Burgett}, J.~S. {Morgan}, J.~L.
  {Tonry}, K.~W. {Hodapp}, S.~{Chastel}, S.~{Chesley}, A.~{Fitzsimmons},
  M.~{Holman}, T.~{Spahr}, D.~{Tholen}, G.~V. {Williams}, S.~{Abe}, J.~D.
  {Armstrong}, T.~H. {Bressi}, R.~{Holmes}, T.~{Lister}, R.~S. {McMillan},
  M.~{Micheli}, E.~V. {Ryan}, W.~H. {Ryan},\ and J.~V. {Scotti} 2013.
\newblock {The Pan-STARRS Moving Object Processing System}.
\newblock {\em \pasp\/}~{\bf 125}, 357--395.

\bibitem[{Dohnanyi}(1969){Dohnanyi}]{Dohnanyi1969}
{Dohnanyi}, J.~S. 1969.
\newblock {Collisional Model of Asteroids and Their Debris}.
\newblock {\em \jgr\/}~{\bf 74}, 2531.

\bibitem[{Dohnanyi}(1971){Dohnanyi}]{Dohnanyi1971}
{Dohnanyi}, J.~S. 1971.
\newblock {Fragmentation and Distribution of Asteroids}.
\newblock {\em NASA Special Publication\/}~{\bf 267}, 263.

\bibitem[{Durda} {\em et~al.}(1998){Durda}, {Greenberg}, and
  {Jedicke}]{Durda1998}
{Durda}, D.~D., R.~{Greenberg},\ and R.~{Jedicke} 1998.
\newblock {A New Interpretation of the Size Distribution of Main-Belt
  Asteroids}.
\newblock In {\em Lunar and Planetary Institute Science Conference Abstracts},
  Volume~29 of {\em Lunar and Planetary Institute Science Conference
  Abstracts}, pp.\  1680.

\bibitem[{Farinella} {\em et~al.}(1998){Farinella}, {Vokrouhlick{\'y}}, and
  {Hartmann}]{Farinella1998}
{Farinella}, P., D.~{Vokrouhlick{\'y}},\ and W.~K. {Hartmann} 1998.
\newblock {Meteorite Delivery via Yarkovsky Orbital Drift}.
\newblock {\em \icarus\/}~{\bf 132}, 378--387.

\bibitem[{Fink} and {Rubin}(2012){Fink} and {Rubin}]{Fink2012}
{Fink}, U.,\ and M.~{Rubin} 2012.
\newblock {The calculation of Af{$\rho$} and mass loss rate for comets}.
\newblock {\em \icarus\/}~{\bf 221}, 721--734.

\bibitem[{Finson} and {Probstein}(1968){Finson} and {Probstein}]{Finson1968}
{Finson}, M.~J.,\ and R.~F. {Probstein} 1968.
\newblock {A theory of dust comets. I. Model and equations}.
\newblock {\em \apj\/}~{\bf 154}, 327--352.

\bibitem[{Gladman} {\em et~al.}(2009){Gladman}, {Davis}, {Neese}, {Jedicke},
  {Williams}, {Kavelaars}, {Petit}, {Scholl}, {Holman}, {Warrington},
  {Esquerdo}, and {Tricarico}]{Gladman2009}
{Gladman}, B.~J., D.~R. {Davis}, C.~{Neese}, R.~{Jedicke}, G.~{Williams}, J.~J.
  {Kavelaars}, J.~{Petit}, H.~{Scholl}, M.~{Holman}, B.~{Warrington},
  G.~{Esquerdo},\ and P.~{Tricarico} 2009.
\newblock {On the asteroid belt's orbital and size distribution}.
\newblock {\em Icarus\/}~{\bf 202}, 104--118.

\bibitem[{Granvik} {\em et~al.}(2009){Granvik}, {Virtanen}, {Oszkiewicz}, and
  {Muinonen}]{Granvik2009}
{Granvik}, M., J.~{Virtanen}, D.~{Oszkiewicz},\ and K.~{Muinonen} 2009.
\newblock {OpenOrb: Open-source asteroid orbit computation software including
  statistical ranging}.
\newblock {\em Meteoritics and Planetary Science\/}~{\bf 44}, 1853--1861.

\bibitem[{Grav} {\em et~al.}(2011){Grav}, {Jedicke}, {Denneau}, {Chesley},
  {Holman}, and {Spahr}]{Grav2011}
{Grav}, T., R.~{Jedicke}, L.~{Denneau}, S.~{Chesley}, M.~J. {Holman},\ and
  T.~B. {Spahr} 2011.
\newblock {The Pan-STARRS Synthetic Solar System Model: A Tool for Testing and
  Efficiency Determination of the Moving Object Processing System}.
\newblock {\em \pasp\/}~{\bf 123}, 423--447.

\bibitem[{Greenberg} {\em et~al.}(1978){Greenberg}, {Hartmann}, {Chapman}, and
  {Wacker}]{Greenberg1978}
{Greenberg}, R., W.~K. {Hartmann}, C.~R. {Chapman},\ and J.~F. {Wacker} 1978.
\newblock {Planetesimals to planets - Numerical simulation of collisional
  evolution}.
\newblock {\em \icarus\/}~{\bf 35}, 1--26.

\bibitem[{Harris}(1996){Harris}]{Harris1996}
{Harris}, A.~W. 1996.
\newblock {The Rotation Rates of Very Small Asteroids: Evidence for 'Rubble
  Pile' Structure}.
\newblock In {\em Lunar and Planetary Science Conference}, Volume~27 of {\em
  Lunar and Planetary Science Conference}, pp.\  493.

\bibitem[{Hodapp} {\em et~al.}(2004){Hodapp}, {Kaiser}, {Aussel}, {Burgett},
  {Chambers}, {Chun}, {Dombeck}, {Douglas}, {Hafner}, {Heasley}, {Hoblitt},
  {Hude}, {Isani}, {Jedicke}, {Jewitt}, {Laux}, {Luppino}, {Lupton}, {Maberry},
  {Magnier}, {Mannery}, {Monet}, {Morgan}, {Onaka}, {Price}, {Ryan},
  {Siegmund}, {Szapudi}, {Tonry}, {Wainscoat}, and {Waterson}]{Hodapp2004}
{Hodapp}, K.~W., N.~{Kaiser}, H.~{Aussel}, W.~{Burgett}, K.~C. {Chambers},
  M.~{Chun}, T.~{Dombeck}, A.~{Douglas}, D.~{Hafner}, J.~{Heasley},
  J.~{Hoblitt}, C.~{Hude}, S.~{Isani}, R.~{Jedicke}, D.~{Jewitt}, U.~{Laux},
  G.~A. {Luppino}, R.~{Lupton}, M.~{Maberry}, E.~{Magnier}, E.~{Mannery},
  D.~{Monet}, J.~{Morgan}, P.~{Onaka}, P.~{Price}, A.~{Ryan}, W.~{Siegmund},
  I.~{Szapudi}, J.~{Tonry}, R.~{Wainscoat},\ and M.~{Waterson} 2004.
\newblock {Design of the Pan-STARRS telescopes}.
\newblock {\em Astronomische Nachrichten\/}~{\bf 325}, 636--642.

\bibitem[{Hsieh}(2009){Hsieh}]{Hsieh2009}
{Hsieh}, H.~H. 2009.
\newblock {The Hawaii trails project: comet-hunting in the main asteroid belt}.
\newblock {\em \aap\/}~{\bf 505}, 1297--1310.

\bibitem[{Hsieh} and {Jewitt}(2006){Hsieh} and {Jewitt}]{Hsieh2006}
{Hsieh}, H.~H.,\ and D.~{Jewitt} 2006.
\newblock {A Population of Comets in the Main Asteroid Belt}.
\newblock {\em Science\/}~{\bf 312}, 561--563.

\bibitem[{Ishiguro} {\em et~al.}(2011){Ishiguro}, {Hanayama}, {Hasegawa},
  {Sarugaku}, {Watanabe}, {Fujiwara}, {Terada}, {Hsieh}, {Vaubaillon}, {Kawai},
  {Yanagisawa}, {Kuroda}, {Miyaji}, {Fukushima}, {Ohta}, {Hamanowa}, {Kim},
  {Pyo}, and {Nakamura}]{Ishiguro2011}
{Ishiguro}, M., H.~{Hanayama}, S.~{Hasegawa}, Y.~{Sarugaku}, J.-i. {Watanabe},
  H.~{Fujiwara}, H.~{Terada}, H.~H. {Hsieh}, J.~J. {Vaubaillon}, N.~{Kawai},
  K.~{Yanagisawa}, D.~{Kuroda}, T.~{Miyaji}, H.~{Fukushima}, K.~{Ohta},
  H.~{Hamanowa}, J.~{Kim}, J.~{Pyo},\ and A.~M. {Nakamura} 2011.
\newblock {Observational Evidence for an Impact on the Main-belt Asteroid (596)
  Scheila}.
\newblock {\em \apjl\/}~{\bf 740}, L11.

\bibitem[{Jacobson} {\em et~al.}(2014){Jacobson}, {Marzari}, {Rossi},
  {Scheeres}, and {Davis}]{Jacobson2014}
{Jacobson}, S.~A., F.~{Marzari}, A.~{Rossi}, D.~J. {Scheeres},\ and D.~R.
  {Davis} 2014.
\newblock {Effect of rotational disruption on the size-frequency distribution
  of the Main Belt asteroid population}.
\newblock {\em \mnras\/}~{\bf 439}, L95--L99.

\bibitem[{Jedicke}(1996){Jedicke}]{Jedicke1996}
{Jedicke}, R. 1996.
\newblock {Detection of Near Earth Asteroids Based Upon Their Rates of Motion}.
\newblock {\em \aj\/}~{\bf 111}, 970--+.

\bibitem[{Jedicke} {\em et~al.}(2002){Jedicke}, {Larsen}, and
  {Spahr}]{Jedicke2002}
{Jedicke}, R., J.~{Larsen},\ and T.~{Spahr} 2002.
\newblock {Observational Selection Effects in Asteroid Surveys}.
\newblock {\em Asteroids III\/}, 71--87.

\bibitem[{Jewitt}(2009){Jewitt}]{Jewitt2009B}
{Jewitt}, D. 2009.
\newblock {The Active Centaurs}.
\newblock {\em \aj\/}~{\bf 137}, 4296--4312.

\bibitem[{Jewitt}(2012){Jewitt}]{Jewitt2012}
{Jewitt}, D. 2012.
\newblock {The Active Asteroids}.
\newblock {\em \aj\/}~{\bf 143}, 66.

\bibitem[{Jewitt} {\em et~al.}(2014){Jewitt}, {Agarwal}, {Li}, {Weaver},
  {Mutchler}, and {Larson}]{Jewitt2014}
{Jewitt}, D., J.~{Agarwal}, J.~{Li}, H.~{Weaver}, M.~{Mutchler},\ and
  S.~{Larson} 2014.
\newblock {Disintegrating Asteroid P/2013 R3}.
\newblock {\em \apjl\/}~{\bf 784}, L8.

\bibitem[{Jewitt} {\em et~al.}(2013){Jewitt}, {Agarwal}, {Weaver}, {Mutchler},
  and {Larson}]{Jewitt2013b}
{Jewitt}, D., J.~{Agarwal}, H.~{Weaver}, M.~{Mutchler},\ and S.~{Larson} 2013.
\newblock {The Extraordinary Multi-tailed Main-belt Comet P/2013 P5}.
\newblock {\em \apjl\/}~{\bf 778}, L21.

\bibitem[{Jewitt} {\em et~al.}(2010){Jewitt}, {Weaver}, {Agarwal}, {Mutchler},
  and {Drahus}]{Jewitt2010}
{Jewitt}, D., H.~{Weaver}, J.~{Agarwal}, M.~{Mutchler},\ and M.~{Drahus} 2010.
\newblock {Newly Disrupted Main Belt Asteroid P/2010 A2}.
\newblock {\em ArXiv e-prints\/}.

\bibitem[{Jewitt} {\em et~al.}(2011){Jewitt}, {Weaver}, {Mutchler}, {Larson},
  and {Agarwal}]{Jewitt2011}
{Jewitt}, D., H.~{Weaver}, M.~{Mutchler}, S.~{Larson},\ and J.~{Agarwal} 2011.
\newblock {Hubble Space Telescope Observations of Main-belt Comet (596)
  Scheila}.
\newblock {\em \apjl\/}~{\bf 733}, L4.

\bibitem[{Kaiser}(2004){Kaiser}]{Kaiser2004}
{Kaiser}, N. 2004.
\newblock {Pan-STARRS: a wide-field optical survey telescope array}.
\newblock In J.~M. {Oschmann}, Jr. (Ed.), {\em Society of Photo-Optical
  Instrumentation Engineers (SPIE) Conference Series}, Volume 5489 of {\em
  Society of Photo-Optical Instrumentation Engineers (SPIE) Conference Series},
  pp.\  11--22.

\bibitem[{Kaiser} {\em et~al.}(2002){Kaiser}, {Aussel}, {Burke}, {Boesgaard},
  {Chambers}, {Chun}, {Heasley}, {Hodapp}, {Hunt}, {Jedicke}, {Jewitt},
  {Kudritzki}, {Luppino}, {Maberry}, {Magnier}, {Monet}, {Onaka}, {Pickles},
  {Rhoads}, {Simon}, {Szalay}, {Szapudi}, {Tholen}, {Tonry}, {Waterson}, and
  {Wick}]{Kaiser2002}
{Kaiser}, N., H.~{Aussel}, B.~E. {Burke}, H.~{Boesgaard}, K.~{Chambers}, M.~R.
  {Chun}, J.~N. {Heasley}, K.-W. {Hodapp}, B.~{Hunt}, R.~{Jedicke},
  D.~{Jewitt}, R.~{Kudritzki}, G.~A. {Luppino}, M.~{Maberry}, E.~{Magnier},
  D.~G. {Monet}, P.~M. {Onaka}, A.~J. {Pickles}, P.~H.~H. {Rhoads}, T.~{Simon},
  A.~{Szalay}, I.~{Szapudi}, D.~J. {Tholen}, J.~L. {Tonry}, M.~{Waterson},\ and
  J.~{Wick} 2002.
\newblock {Pan-STARRS: A Large Synoptic Survey Telescope Array}.
\newblock In J.~A. {Tyson} and S.~{Wolff} (Eds.), {\em Society of Photo-Optical
  Instrumentation Engineers (SPIE) Conference Series}, Volume 4836 of {\em
  Society of Photo-Optical Instrumentation Engineers (SPIE) Conference Series},
  pp.\  154--164.

\bibitem[{Kantor}(2013){Kantor}]{Kantor2013}
{Kantor}, J. 2013.
\newblock personal communication.

\bibitem[{Kubica} {\em et~al.}(2007){Kubica}, {Denneau}, {Grav}, {Heasley},
  {Jedicke}, {Masiero}, {Milani}, {Moore}, {Tholen}, and
  {Wainscoat}]{Kubica2007}
{Kubica}, J., L.~{Denneau}, T.~{Grav}, J.~{Heasley}, R.~{Jedicke},
  J.~{Masiero}, A.~{Milani}, A.~{Moore}, D.~{Tholen},\ and R.~J. {Wainscoat}
  2007.
\newblock {Efficient intra- and inter-night linking of asteroid detections
  using kd-trees}.
\newblock {\em \icarus\/}~{\bf 189}, 151--168.

\bibitem[{Larson}(2007){Larson}]{Larson2007}
{Larson}, S. 2007.
\newblock {Current NEO surveys}.
\newblock In G.~B. {Valsecchi} and D.~{Vokrouhlick{\'y}} (Eds.), {\em IAU
  Symposium}, Volume 236 of {\em IAU Symposium}, pp.\  323--328.

\bibitem[{Larson} {\em et~al.}(1998){Larson}, {Brownlee}, {Hergenrother}, and
  {Spahr}]{Larson1998}
{Larson}, S., J.~{Brownlee}, C.~{Hergenrother},\ and T.~{Spahr} 1998.
\newblock {The Catalina Sky Survey for NEOs}.
\newblock In {\em Bulletin of the American Astronomical Society}, Volume~30 of
  {\em Bulletin of the American Astronomical Society}, pp.\  1037.

\bibitem[{Magnier}(2006){Magnier}]{Magnier2006}
{Magnier}, E. 2006.
\newblock {The Pan-STARRS PS1 Image Processing Pipeline}.
\newblock In {\em The Advanced Maui Optical and Space Surveillance Technologies
  Conference}.

\bibitem[{Michel} {\em et~al.}(2002){Michel}, {Tanga}, {Benz}, and
  {Richardson}]{Michel2002}
{Michel}, P., P.~{Tanga}, W.~{Benz},\ and D.~C. {Richardson} 2002.
\newblock {Formation of Asteroid Families by Catastrophic Disruption:
  Simulations with Fragmentation and Gravitational Reaccumulation}.
\newblock {\em \icarus\/}~{\bf 160}, 10--23.

\bibitem[{Milani} {\em et~al.}(2012){Milani}, {Kne{\v z}evi{\'c}},
  {Farnocchia}, {Bernardi}, {Jedicke}, {Denneau}, {Wainscoat}, {Burgett},
  {Grav}, {Kaiser}, {Magnier}, and {Price}]{Milani2012}
{Milani}, A., Z.~{Kne{\v z}evi{\'c}}, D.~{Farnocchia}, F.~{Bernardi},
  R.~{Jedicke}, L.~{Denneau}, R.~J. {Wainscoat}, W.~{Burgett}, T.~{Grav},
  N.~{Kaiser}, E.~{Magnier},\ and P.~A. {Price} 2012.
\newblock {Identification of known objects in Solar System surveys}.
\newblock {\em \icarus\/}~{\bf 220}, 114--123.

\bibitem[{Milani} and {Knezevic}(1994){Milani} and {Knezevic}]{Milani1994}
{Milani}, A.,\ and Z.~{Knezevic} 1994.
\newblock {Asteroid proper elements and the dynamical structure of the asteroid
  main belt}.
\newblock {\em \icarus\/}~{\bf 107}, 219--254.

\bibitem[{Muinonen} {\em et~al.}(2006){Muinonen}, {Virtanen}, {Granvik}, and
  {Laakso}]{Muinonen2006}
{Muinonen}, K., J.~{Virtanen}, M.~{Granvik},\ and T.~{Laakso} 2006.
\newblock {Asteroid orbits using phase-space volumes of variation}.
\newblock {\em \mnras\/}~{\bf 368}, 809--818.

\bibitem[{Novakovic} {\em et~al.}(2014){Novakovic}, {Hsieh}, {Cellino},
  {Micheli}, and {Pedani}]{Novakovic2014}
{Novakovic}, B., H.~H. {Hsieh}, A.~{Cellino}, M.~{Micheli},\ and M.~{Pedani}
  2014.
\newblock {Discovery of a young asteroid cluster associated with P/2012~F5
  (Gibbs)}.
\newblock {\em ArXiv e-prints\/}.

\bibitem[{O'Brien}(2009){O'Brien}]{Obrien2009}
{O'Brien}, D.~P. 2009.
\newblock {The Yarkovsky effect is not responsible for small crater depletion
  on Eros and Itokawa}.
\newblock {\em \icarus\/}~{\bf 203}, 112--118.

\bibitem[{O'Brien} and {Greenberg}(2003){O'Brien} and {Greenberg}]{OBrien2003}
{O'Brien}, D.~P.,\ and R.~{Greenberg} 2003.
\newblock {Steady-state size distributions for collisional populations:.
  analytical solution with size-dependent strength}.
\newblock {\em \icarus\/}~{\bf 164}, 334--345.

\bibitem[{Parker} {\em et~al.}(2008){Parker}, {Ivezi{\'c}}, {Juri{\'c}},
  {Lupton}, {Sekora}, and {Kowalski}]{Parker2008}
{Parker}, A., {\v Z}.~{Ivezi{\'c}}, M.~{Juri{\'c}}, R.~{Lupton}, M.~D.
  {Sekora},\ and A.~{Kowalski} 2008.
\newblock {The size distributions of asteroid families in the SDSS Moving
  Object Catalog 4}.
\newblock {\em \icarus\/}~{\bf 198}, 138--155.

\bibitem[{Pravec} {\em et~al.}(2012){Pravec}, {Harris}, {Ku{\v s}nir{\'a}k},
  {Gal{\'a}d}, and {Hornoch}]{Pravec2012}
{Pravec}, P., A.~W. {Harris}, P.~{Ku{\v s}nir{\'a}k}, A.~{Gal{\'a}d},\ and
  K.~{Hornoch} 2012.
\newblock {Absolute magnitudes of asteroids and a revision of asteroid albedo
  estimates from WISE thermal observations}.
\newblock {\em \icarus\/}~{\bf 221}, 365--387.

\bibitem[{Reach} {\em et~al.}(2010){Reach}, {Vaubaillon}, {Lisse}, {Holloway},
  and {Rho}]{Reach2010}
{Reach}, W.~T., J.~{Vaubaillon}, C.~M. {Lisse}, M.~{Holloway},\ and J.~{Rho}
  2010.
\newblock {Explosion of Comet 17P/Holmes as revealed by the Spitzer Space
  Telescope}.
\newblock {\em \icarus\/}~{\bf 208}, 276--292.

\bibitem[{S{\'a}nchez} and {Scheeres}(2013){S{\'a}nchez} and
  {Scheeres}]{Sanchez2013}
{S{\'a}nchez}, P.,\ and D.~J. {Scheeres} 2013.
\newblock {The Strength of Regolith and Rubble Pile Asteroids}.
\newblock {\em ArXiv e-prints\/}.

\bibitem[{Scheeres} and {Sanchez}(2012){Scheeres} and {Sanchez}]{Scheeres2012}
{Scheeres}, D.~J.,\ and P.~{Sanchez} 2012.
\newblock {The Strength of Rubble Pile Asteroids}.
\newblock {\em AGU Fall Meeting Abstracts\/}, A5.

\bibitem[{Shah} {\em et~al.}(2013){Shah}, {Woods}, {Faccenda}, {Johnson},
  {Lambour}, {Pearce}, and {Stuart}]{Shah2013}
{Shah}, R., D.~F. {Woods}, W.~{Faccenda}, J.~{Johnson}, R.~{Lambour}, E.~C.
  {Pearce},\ and J.~S. {Stuart} 2013.
\newblock "asteroid detection with the space surveillance telescope".
\newblock In {\em AMOS Conference Technical Papers}.

\bibitem[{Snodgrass} {\em et~al.}(2010){Snodgrass}, {Tubiana}, {Vincent},
  {Sierks}, {Hviid}, {Moissi}, {Boehnhardt}, {Barbieri}, {Koschny}, {Lamy},
  {Rickman}, {Rodrigo}, {Carry}, {Lowry}, {Laird}, {Weissman}, {Fitzsimmons},
  {Marchi}, and {OSIRIS Team}]{Snodgrass2010}
{Snodgrass}, C., C.~{Tubiana}, J.-B. {Vincent}, H.~{Sierks}, S.~{Hviid},
  R.~{Moissi}, H.~{Boehnhardt}, C.~{Barbieri}, D.~{Koschny}, P.~{Lamy},
  H.~{Rickman}, R.~{Rodrigo}, B.~{Carry}, S.~C. {Lowry}, R.~J.~M. {Laird},
  P.~R. {Weissman}, A.~{Fitzsimmons}, S.~{Marchi},\ and {OSIRIS Team} 2010.
\newblock {A collision in 2009 as the origin of the debris trail of asteroid
  P/2010A2}.
\newblock {\em \nat\/}~{\bf 467}, 814--816.

\bibitem[{Stevenson} and {Jewitt}(2012){Stevenson} and
  {Jewitt}]{Stevenson2012a}
{Stevenson}, R.,\ and D.~{Jewitt} 2012.
\newblock {Near-nucleus Photometry of Outbursting Comet 17P/Holmes}.
\newblock {\em \aj\/}~{\bf 144}, 138.

\bibitem[{Stevenson} {\em et~al.}(2012){Stevenson}, {Kramer}, {Bauer},
  {Masiero}, and {Mainzer}]{Stevenson2012b}
{Stevenson}, R., E.~A. {Kramer}, J.~M. {Bauer}, J.~R. {Masiero},\ and A.~K.
  {Mainzer} 2012.
\newblock {Characterization of Active Main Belt Object P/2012 F5 (Gibbs): A
  Possible Impacted Asteroid}.
\newblock {\em \apj\/}~{\bf 759}, 142.

\bibitem[{Tonry}(2011){Tonry}]{Tonry2011}
{Tonry}, J.~L. 2011.
\newblock {An Early Warning System for Asteroid Impact}.
\newblock {\em \pasp\/}~{\bf 123}, 58--73.

\bibitem[{Tsuchiyama} {\em et~al.}(2011){Tsuchiyama}, {Uesugi}, {Matsushima},
  {Michikami}, {Kadono}, {Nakamura}, {Uesugi}, {Nakano}, {Sandford}, {Noguchi},
  {Matsumoto}, {Matsuno}, {Nagano}, {Imai}, {Takeuchi}, {Suzuki}, {Ogami},
  {Katagiri}, {Ebihara}, {Ireland}, {Kitajima}, {Nagao}, {Naraoka}, {Noguchi},
  {Okazaki}, {Yurimoto}, {Zolensky}, {Mukai}, {Abe}, {Yada}, {Fujimura},
  {Yoshikawa}, and {Kawaguchi}]{Tsuchiyama2011}
{Tsuchiyama}, A., M.~{Uesugi}, T.~{Matsushima}, T.~{Michikami}, T.~{Kadono},
  T.~{Nakamura}, K.~{Uesugi}, T.~{Nakano}, S.~A. {Sandford}, R.~{Noguchi},
  T.~{Matsumoto}, J.~{Matsuno}, T.~{Nagano}, Y.~{Imai}, A.~{Takeuchi},
  Y.~{Suzuki}, T.~{Ogami}, J.~{Katagiri}, M.~{Ebihara}, T.~R. {Ireland},
  F.~{Kitajima}, K.~{Nagao}, H.~{Naraoka}, T.~{Noguchi}, R.~{Okazaki},
  H.~{Yurimoto}, M.~E. {Zolensky}, T.~{Mukai}, M.~{Abe}, T.~{Yada},
  A.~{Fujimura}, M.~{Yoshikawa},\ and J.~{Kawaguchi} 2011.
\newblock {Three-Dimensional Structure of Hayabusa Samples: Origin and
  Evolution of Itokawa Regolith}.
\newblock {\em Science\/}~{\bf 333}, 1125--.

\bibitem[{Virtanen} {\em et~al.}(2001){Virtanen}, {Muinonen}, and
  {Bowell}]{Virtanen2001}
{Virtanen}, J., K.~{Muinonen},\ and E.~{Bowell} 2001.
\newblock {Statistical Ranging of Asteroid Orbits}.
\newblock {\em \icarus\/}~{\bf 154}, 412--431.

\bibitem[{Walsh} {\em et~al.}(2009){Walsh}, {Michel}, and
  {Richardson}]{Walsh2009}
{Walsh}, K.~J., P.~{Michel},\ and D.~C. {Richardson} 2009.
\newblock {Collisional and Rotational Disruption of Asteroids}.
\newblock {\em ArXiv e-prints\/}.

\bibitem[{Weisstein}(2014){Weisstein}]{SphericalCap}
{Weisstein}, E.~W. 2014.
\newblock {{Spherical Cap} From MathWorld: A Wolfram Web Resource}.

\bibitem[{Willman} {\em et~al.}(2010){Willman}, {Jedicke}, {Moskovitz},
  {Nesvorn{\'y}}, {Vokrouhlick{\'y}}, and {Moth{\'e}-Diniz}]{Willman2010}
{Willman}, M., R.~{Jedicke}, N.~{Moskovitz}, D.~{Nesvorn{\'y}},
  D.~{Vokrouhlick{\'y}},\ and T.~{Moth{\'e}-Diniz} 2010.
\newblock {Using the youngest asteroid clusters to constrain the space
  weathering and gardening rate on S-complex asteroids}.
\newblock {\em \icarus\/}~{\bf 208}, 758--772.

\bibitem[{Zappala} {\em et~al.}(1990){Zappala}, {Cellino}, {Farinella}, and
  {Knezevic}]{Zappala1990}
{Zappala}, V., A.~{Cellino}, P.~{Farinella},\ and Z.~{Knezevic} 1990.
\newblock {Asteroid families. I - Identification by hierarchical clustering and
  reliability assessment}.
\newblock {\em \aj\/}~{\bf 100}, 2030--2046.

\end{thebibliography}
\bibliographystyle{icarus}

\end{document}